\documentclass[%
reprint,
amsmath,amssymb,
prx,
]{revtex4-2}

\usepackage{graphicx}
\usepackage{dcolumn}
\usepackage{bm}

\usepackage{slashed}
\usepackage{xcolor}

\usepackage{amsmath}
\usepackage{amssymb}
\usepackage{mathrsfs}

\usepackage[hidelinks]{hyperref}
\usepackage{url}

\DeclareMathOperator\arctanh{arctanh}
\begin{document}

\title{Nonperturbative $\phi^4$ potentials: Phase transitions and light horizons}

\author{Yuan Shi}
\email{shi9@llnl.gov}
\affiliation{Lawrence Livermore National Laboratory, Livermore, California 94551, USA}

\date{\today}

\begin{abstract}
It is commonly believed that a massive real scalar field $\phi$ only mediates short-range interactions on the scale of its Compton wavelength via the Yukawa potential. However, in the nonperturbative regime of nonlinear self coupling, $\phi$ can also mediate larger scale interactions. Moreover, the classical potential, namely, the static configuration of $\phi$ in the presence of an external source, is not always unique for given boundary conditions. 
In this paper, a complete set of finite-energy potentials (FEPs) induced by a Gaussian source is identified in one, two, and three spatial dimensions when the nonlinearity is of the Mexican-hat type, which is often prescribed to induce spontaneous symmetry breaking. In the size-strength parameter space of the source, phase boundaries are mapped out, across which the number of FEPs differ. Additionally, softer phase transitions are delineated according to whether the potential exhibits a light horizon at which $\phi$ vanishes.
The light horizon is of physical significance when $\phi$ couples with other particles. For example, when $\phi$ is the Higgs field, all elementary particles become massless at the light horizon. It is remarkable that white dwarfs and neutron stars are potentially in a phase where light horizons exist, whose outer radii are a few times the star sizes. Moreover, suppose elementary fermions of mass greater than \mbox{$\sim10^3$ GeV} exist, then they may also be surrounded by light horizons with radii comparable to the Higgs Compton wavelength. 
Finally, nonperturbative states may also be realized in condensed matter systems, wherein phase transitions are controllable using localized sources. 
\end{abstract}

\maketitle

\section{Introduction}
In the presence of nonlinearities, solutions to classical field equations may contain configurations that are distinct from wave-like solutions \cite{Weinberg2012classical}. For example, in one spatial dimension, a self-interacting real scalar field $\phi$ exhibits kink solutions \cite{Dashen74extended}, whose asymptotic values $\phi(+\infty)$ and $\phi(-\infty)$ are different. 
In two spatial dimensions, with a complex scalar field coupling to a U(1) gauge field, localized vortex solutions exist for which the scalar field is depleted at locations where the magnetic flux penetrates \cite{Abrikosov1957}. 
Other nontrivial field configurations exist in higher dimensions when the gauge group is nonabelian, such as monopoles in three spatial dimensions \cite{Hooft1974magnetic,Polyakov74} and instantons in four spatial dimensions \cite{Belavin1975}. 
Nonperturbative field configurations are not topologically equivalent to the trivial vacuum, in the sense that they cannot be obtained by continuously deforming of the vacuum at a finite energy cost. Moreover, configurations that have different topological charges are also disconnected from each other in perturbation theory, so each topologically equivalent class must be included separately in path integrals \cite{Coleman1985}.
Accounting for nonperturbative configurations may lead to additional terms in the effective Lagrangian, which may have profound consequences, including breaking the fundamental symmetries of the field theory \cite{Hooft76symmetry}.

In this paper, I will show that additional nonperturbative field configurations exist in the presence of external sources. External sources arise when one focuses on a subsystem of a bigger problem. 
A familiar example is the Coulomb potential, which acquires a nontrivial configuration when external charges are prescribed. Another example is the Yukawa potential, which arises due to the $\phi\bar{\psi}\psi$ coupling: When the expectation value of fermions $\langle\bar{\psi}\psi\rangle$ is nonzero, the classical field equation for $\phi$ acquires a finite source term.
The significance of an external source is that it adds to the energy integral a term whose sign is indefinite. Consequently, Derrick's theorem \cite{Derrick1964comments}, which uses a scaling argument to show that specific finite-energy configurations only exist in specific spatial dimensions, no longer applies. 
Moreover, the source term also appears in the conservation law of topological charges, so field configurations in different topological classes become connected via the external source. 
Since we are living in a universe that is filled with matter, and possibly dark matter \cite{Arbey2021}, it is important to also consider field configurations in the presence of external sources.

As an example of smooth and localized sources, this paper will focus on an isolated Gaussian. Sources of other profiles, which will be briefly discussed in Sec.~\ref{sec:discussion}, lead to similar features.
On a microscopic scale, the Gaussian source may approximate the wave packet of a single particle; On a macroscopic scale, the Gaussian source may approximate the averaged distribution of an aggregated amount of matter. Regardless of the physical origin of the source term, the resultant spherically symmetric static field configuration in $D$ spatial dimension satisfies
\begin{equation}
\label{eq:EOM_dimensionful}
\frac{\partial^2\phi}{\partial r^2}+\frac{D-1}{r}\frac{\partial\phi}{\partial r} = \frac{\partial U}{\partial\phi} + \mathscr{S},
\end{equation}
where $U(\phi)$ is the self interaction potential and $\mathscr{S}(r)$ is the source term.
The physical significance of static field configurations is that they provide low-energy descriptions of the $\phi$-mediated interaction between a test particle and the external source when their relative motion is nonrelativistic. 
For a Gaussain source, $\mathscr{S}=(\alpha/a^D\pi^{D/2}) \exp(-r^2/a^2)$, 
which becomes a $D$-dimensional $\delta$-function when $a\rightarrow 0$. 
Knowing $\phi$ due to the $\delta$-function source is sufficient for linear theories, in which potentials resultant from other sources can be constructed using convolutions. 
However, in nonlinear theories, the behavior of $\phi$ is qualitatively different for sources of different size $a$ and strength $\alpha$. The goal of this paper is to map out the behaviors of $\phi$ in the source size-strength parameter space.

Apart from the external source, the other key ingredient for nontrivial field configurations is the nonlinearity, which enters via the self interaction potential. 
A particularly important case is the Maxican-hat potential which, up to some shifts and rescalings, can be written as
\begin{equation}
\label{eq:Maxican-hat}
U=\frac{\lambda}{4!}(\phi^2-v^2)^2,
\end{equation}
where the constant $v>0$ is the vacuum expectation value (VEV) and the coupling constant $\lambda>0$ so that the theory is temporally stable. The Maxican-hat potential is often prescribed to model spontaneous symmetry breaking: There exist two degenerate ground states $\phi=\pm v$ in the absence of source terms.  
Spontaneous symmetry breaking plays an important role in field theories. For example, in the phenomenological model of superconductivity \cite{Ginzburg50}, a U(1) version of the spontaneous symmetry breaking emulates effects of Cooper-pair formation below the superconducting transition temperature \cite{BCS57}. 
Moreover, in the standard model of particle physics, an SU(2) version of the spontaneous symmetry breaking is invoked to explain a finite Higgs VEV that endows elementary particles with their masses \cite{Higgs64}.

Given the source and the nonlinearity as two key ingredients, to study classical potentials, namely, static field configurations, it is convenient to nondimensionalized the field equation. In the natural units $\hbar=c=1$, since the action $\int dt d^Dx [\frac{1}{2}(\partial_\mu\phi)^2-U-\mathscr{S}\phi]$ is dimensionless, $\phi$ has the mass dimension $(D-1)/2$. Then, in Eq.~(\ref{eq:Maxican-hat}), $v$ has the same unit as $\phi$, the self coupling $\lambda$ has the mass dimension $3-D$, and the strength of the Gaussian source $\alpha$ has the mass dimension $(3-D)/2$. 
Thus, a natural scheme to nondimensionalize Eq.~(\ref{eq:EOM_dimensionful}) is to define $q=\phi/v, \rho=rm, R=am$, and $A=\alpha m^{D-2}/v$. Here, $m=v\sqrt{\lambda/3}$ is the mass of $\phi$, and $m^2$ is defined as $\partial^2 U/\partial\phi^2$ at the minimum of $U$. 
The static field equation can then be put into a form that is reminiscent of classical dynamical systems in phase space:
\begin{eqnarray}
\label{eq:EOM}
\dot{p} + \frac{D-1}{\rho}p = \frac{1}{2}q(q^2-1)+S_0e^{-\rho^2/R^2},
\end{eqnarray}
where $\dot{p}=dp/d\rho$, $p=dq/d\rho$, and $S_0=A/R^D\pi^{D/2}$.
Notice that the equation is invariant under
$q\rightarrow -q, p\rightarrow -p$, and $A\rightarrow -A$. Therefore, it is sufficient to consider the case $A\ge0$.
Moreover, notice that when $\rho\rightarrow -\rho$ the equation is invariant under $q\rightarrow q$ and $p\rightarrow -p$. Therefore, we can focus on $q(\rho)$ that is an even function. The limit $\rho\rightarrow0$ is well behaved if $p(0)\sim \rho$. 
Alternatively, $q\simeq c\rho^{-\gamma}$. Matching the leading divergence requires $\gamma=1$ and $c^2=2(3-D)$, which is possible when $D<3$.

Among an infinite number of solutions to the static field equation, only a countable number of potentials have finite energy. The energy of a given field configuration is $E=\int d^Dx [\frac{1}{2}(\partial_t\phi)^2 + \frac{1}{2}(\nabla\phi)^2 +U + \mathscr{S}\phi]$, and the time derivative vanishes for static field configurations. For spherically symmetric source and potentials, $E=v^2m^{2-D} \sigma_{D-1} H$, where $\sigma_{D}$ is the surface area of the unit $D$-sphere, and the normalized energy $H=\int_0^\infty d\rho \rho^{D-1}\mathcal{E}$. The normalized energy density is
\begin{equation}
\label{eq:energy_density}
\mathcal{E} = \frac{1}{2}p^2+\frac{1}{8}(q^2-1)^2+S q,
\end{equation}
where $S=S_0\exp(-\rho^2/R^2)$. 
The extremum of $H$, which satisfies $\delta H/\delta q=0$, leads to Eq.~(\ref{eq:EOM}). 
To have a finite energy, the potential must be well behaved near the origin, which is possible because the smoothness of the source regularizes the behavior of $q$ at $\rho\sim0$.
Then, a finite energy only requires that $\mathcal{E}$ approaches zero faster than $\rho^{-D}$ when $\rho\rightarrow\infty$. Since $S$ falls off rapidly, the energy is finite as long as $q$ approaches $\pm1$ faster than $\rho^{-D/2}$. 
In fact, when the source term vanishes and $q\rightarrow\pm1$, the potentials becomes the Yukawa potential, which falls off as $\sim\exp(-\rho)/\rho^{(D-1)/2}$ when $\rho\rightarrow\infty$. Therefore, the spherically symmetric potential is of finite energy if and only if $p(0)=0$ and $q(\infty)=\pm1$.

For a given boundary condition, the existence and uniqueness of finite-energy potentials (FEPs) are not guaranteed. To see why, notice that as a classical dynamical system, Eq.~(\ref{eq:EOM}) has three fixed points in the $qp$-phase space at $(0,0)$ and $(\pm1,0)$ when $S_0=0$. While $(0,0)$ is a stable fixed point, $(\pm1,0)$ are unstable saddle points. 
In other words, the boundary conditions $q(\infty)=\pm1$, which may lead to FEPs that are stable against temporal fluctuations, are associated with solutions that are unpredictable against spatial uncertainties.
This intriguing situation, although perhaps not surprising, may be better appreciated through the following thought experiment: Suppose an experimentalist measures the Higgs expectation value $\langle\phi\rangle$ to be $v_1$ near the sea level of Earth, which serves as a localized source; and another experimentalist, who lives on the top of a mountain, measures $\langle\phi\rangle$ to be $v_2$. Then, if there is any uncertainty in the measurements, one cannot reliably predict what $\langle\phi\rangle$ is at the edge of the solar system, without assuming that the classical field is in a finite-energy configuration.

As will be shown in this paper, it turns out that exactly how many different FEPs exist depends on both the size $R$ and the strength $A$ of the source, as well as the dimensionality $D$ of the problem. 
Due to symmetries of Eq.~(\ref{eq:EOM}), we can assume without loss of generality that $A\ge0$. Then, there always exist a unique ground state configuration with $q(\infty)=-1$. The situation is more subtle for $q(\infty)=1$, which corresponds to excited states.
For example, in one spatial dimension (1D), there exist two FEPs with $q(\infty)=1$ when $A<A_c$, while no FEP with $q(\infty)=1$ when $A>A_c$ (Sec.~\ref{sec:1D}). The critical value $A_c\simeq1$ for $R\ll1$; and for $R\gg1$, $A_c$ is such that $S_0\simeq1/3\sqrt{3}$.
The behavior is different in higher spatial dimensions (Sec.~\ref{sec:DD}): When $R\ll1$ there always exist a unique FEP with $q(\infty)=1$; When $R\gg1$, there is also a unique FEP with $q(\infty)=1$ if $A<A_{c1}$ or $A>A_{c2}$. However, when $A_{c1}<A<A_{c2}$, there exist three different FEPs with $q(\infty)=1$.

The $R$-$A$ parameter space can thus be divided into different phases, which will be mapped out completely in this paper by combining asymptotic and numerical solutions. 
As an informative name, each phase will be labeled by a four-digit number $N_+^+N_-^+N_-^-N_+^-$, where $N_+^+$ is the number of FEPs with $q(0)>0$ and $q(\infty)=1$, $N_-^+$ is the number of FEPs with $q(0)<0$ and $q(\infty)=1$, and so on. Obviously, the sum of four digits is the total number of FEPs in that phase.
For example, in 1D, the region $R\ll1$ and $A<1$ is the 1110 phase, while the region $R\ll1$ and $A>1$ is the 0010 phase. 
In addition to relying on asymptotic solutions for $R\gg1$ or $R\ll1$, numerical solutions are used to map out the phase boundaries when $R\sim 1$. 
A nominal second-order algorithm is constructed using a central difference scheme by approximating $\ddot{q}(\rho_n)\simeq(q_{n+1}-2q_n+q_{n-1})/h^2$ and $\dot{q}(\rho_n)\simeq(q_{n+1}-q_{n-1})/(2h)$, where $q_n=q(\rho_n)$, $\rho_n=nh$ for $n\in\mathbb{N}$, and $h$ is the step size. 
The step size is taken to be $h=\varepsilon$ min$(1,R)$, which resolves both the intrinsic scale of the nonlinear problem $\rho\sim 1$, and the scale length of the source $\rho\sim R$. 
At any finite $h$, numerical convergence is inevitably poor near the saddle points. 
Nevertheless, away from the saddle points, numerical results are hardly distinguishable for $\varepsilon=10^{-3}$ and $\varepsilon=10^{-2}$, the later of which is used for all numerical results reported in this paper. 
For given $R$ and $A$ values, initial conditions $q(0)$ and $p(0)$ are scanned to search for boundaries across which numerical solutions transition from oscillatory to divergent, and these phase-space boundaries correspond to FEPs.
Subsequently, the source parameters $R$ and $A$ are scanned to determine parameter-space boundaries across which FEPs exhibit distinct phases.

In addition to discrete jumps of the number of FEPs, there also exist softer phase transitions, which can be characterized using the radius of light horizons as an order parameter. A light horizon is a sphere in $D$-dimensional space where $q(\rho_c)=0$, which is a generalization of domain walls for kink solutions. 
To see how light horizons emerge, naively, $q(\infty)=1$ is slightly depleted by a weak source; and for a stronger source, the VEV is depleted more. A light horizon emerges when the source exceeds a critical strength, beyond which $q$ crosses zero. 
When $\phi$ is the Higgs field, all elementary particles become massless at the light horizon where they move at the speed of light, therefore the name ``light horizon".
Since solutions to Eq.~(\ref{eq:EOM}) are smooth, $q^2$ is noticeably depleted in a thin spherical shell surrounding $\rho_c$, whose thickness $\Delta\rho$ is at least the Compton wavelength. The exact values of $\rho_c$ and $\Delta\rho$, and whether the light horizon exists at all, again depend on $R$ and $A$.

As a nonperturbative effect, the light horizon may have a radius that is much larger than the Compton wavelength.
For example (Sec.~\ref{sec:DD}), when $D>1$ and $R\gg1$, two of the four FEPs exhibit light horizons when the source strength satisfies $(D-1)\sqrt{2e}/(3R) <S_0 \lesssim 1/3\sqrt{3}$, where $e$ is the base of natural logarithm. In this 1210 phase, the larger light horizon has a radius of $\rho_c\simeq R\sqrt{\ln[(3RS_0)/(D-1)]}$ and the smaller light horizon has a radius of $\rho_c\simeq (D-1)/(3S_0)$. 
Notice that $\rho_c$ is potentially macroscopic. 
Moreover, it is worth noting that the critical source density for reaching the 1210 phase decreases with $R$ (Sec.~\ref{sec:discussion}). For a source of size \mbox{$a\sim1$ m}, the critical mass density \mbox{$\sim10^{14}\; \text{kg/m}^3$} is enormous for the Higgs field in 3D, 
so usual objects in our daily life are well within the perturbative phase. 
On the other hand, when \mbox{$a\sim10^6$ km}, which is on the order of the solar radius, the critical mass density is reduced to \mbox{$\sim10^5\; \text{kg/m}^3$}, which suggests that stars are not far away from reaching the nonperturbative phase. Indeed, white dwarfs and neutron stars are likely within the 1210 phase where light horizons exist, whose outer radii are typically a few times the star sizes.
In Sec.~\ref{sec:discussion}, physical implications of nonperturbative FEPs will be discussed both for the Higgs field and other effective fields.

\section{Potentials in one dimension \label{sec:1D}}
Building upon the kink solutions, let us first consider potentials in 1D.   
Since the singular term $p/\rho$ is absent from Eq.~(\ref{eq:EOM}), 
1D potentials are special. Nevertheless, many features of 1D potentials are preserved in higher dimensions. 
In particular, notice that when $\rho\rightarrow\infty$, the source term vanishes and Eq.~(\ref{eq:EOM}) approaches $\ddot{q}=q(q^2-1)/2$, which is integrable. 
Therefore, no chaotic behavior is expected for a localized source term, and static potentials are asymptotic to the 1D case when $\rho\rightarrow\infty$.
The general solutions to the 1D source-free problem are given by Jacobi elliptic functions, which are periodic and possibly divergent. Except for the special case where the period of elliptic functions is infinite, which corresponds to the kink solutions, the total energy of classical potentials are infinite.

Key results that will be shown in this section are summarized here: (i) When $R\ll1$, the system is in the 1110 phase when $A<A_{1110}^{0010}$ and in the 0010 phase when $A>A_{1110}^{0010}$. The phase boundary is 
\begin{equation}
\label{eq:A_1110^0010}
A_{1110}^{0010} \simeq 1,
\end{equation} 
which is independent of R. In the 1110 phase, there exists one FEP that exhibits a light horizon with radius
\begin{equation}
\label{eq:rc_1DRs}
\rho_c\simeq2\arctanh\sqrt{1-A},
\end{equation}
which diverges as $\rho_c\sim\ln(4/A)$ when $A\rightarrow 0$ and terminates as $\rho_c\sim 2\sqrt{1-A}$ when $A$ increases towards 1. 
(ii) When $R\gg1$, the system transitions through the 1110, 2010, and 0010 phases as $A$ increases. The phase boundaries are
\begin{equation}
\label{eq:A_1110^2010}
A_{1110}^{2010} \simeq \sqrt{\frac{\pi}{6}}\frac{R}{3},\quad A_{2010}^{0010}\simeq \sqrt{\frac{\pi}{3}}\frac{R}{3}.
\end{equation}
In the 1110 phase, there exists one nonperturbative FEP that exhibits a light horizon with radius
\begin{equation}
\label{eq:rc_1DRl}
\rho_c\simeq\frac{1}{k} \text{arcsinh} \sqrt{-\frac{(2+\theta)q_c}{2q_\infty}},
\end{equation}
which diverges as $\rho_c\sim \ln(2/\sqrt{S_0})$ when $S_0\rightarrow 0$ and terminates as $\rho_c\sim6^{5/4}\sqrt{S_c-S_0}$ when $S_0$ increases towards the 1110-2010 phase boundary $S_c=1/3\sqrt{6}$. In the above formula, $q_\infty$ is the largest positive root of $\frac{1}{2}q_\infty(q_\infty^2-1)+S_0=0$, $q_c=-q_\infty+f<0$ is the critical initial value for the nonperturbative FEP, $f=\sqrt{2(1-q_\infty^2)}$, and $\theta=(q_\infty-q_c)/(q_\infty+q_c)$. The thickness of the light horizon is $\Delta\rho\sim1/k$, where $k=\sqrt{(3q_\infty^2-1)/8}>1/2\sqrt{2}$ in the 1110 phase where the light horizon exists. 
(iii) Numerical results are summarized in Fig.~\ref{fig:1D}, which shows no additional phase when $R\sim1$. In other words, the 1D problem has three phases in total, and one FEP has a light horizon, whose radius diverges logarithmically when the source strength diminishes.

\subsection{Asymptotic solutions for $R\ll1$ \label{sec:1D:delta}}
\begin{figure}[b]
	\centering
	\includegraphics[width=0.48\textwidth]{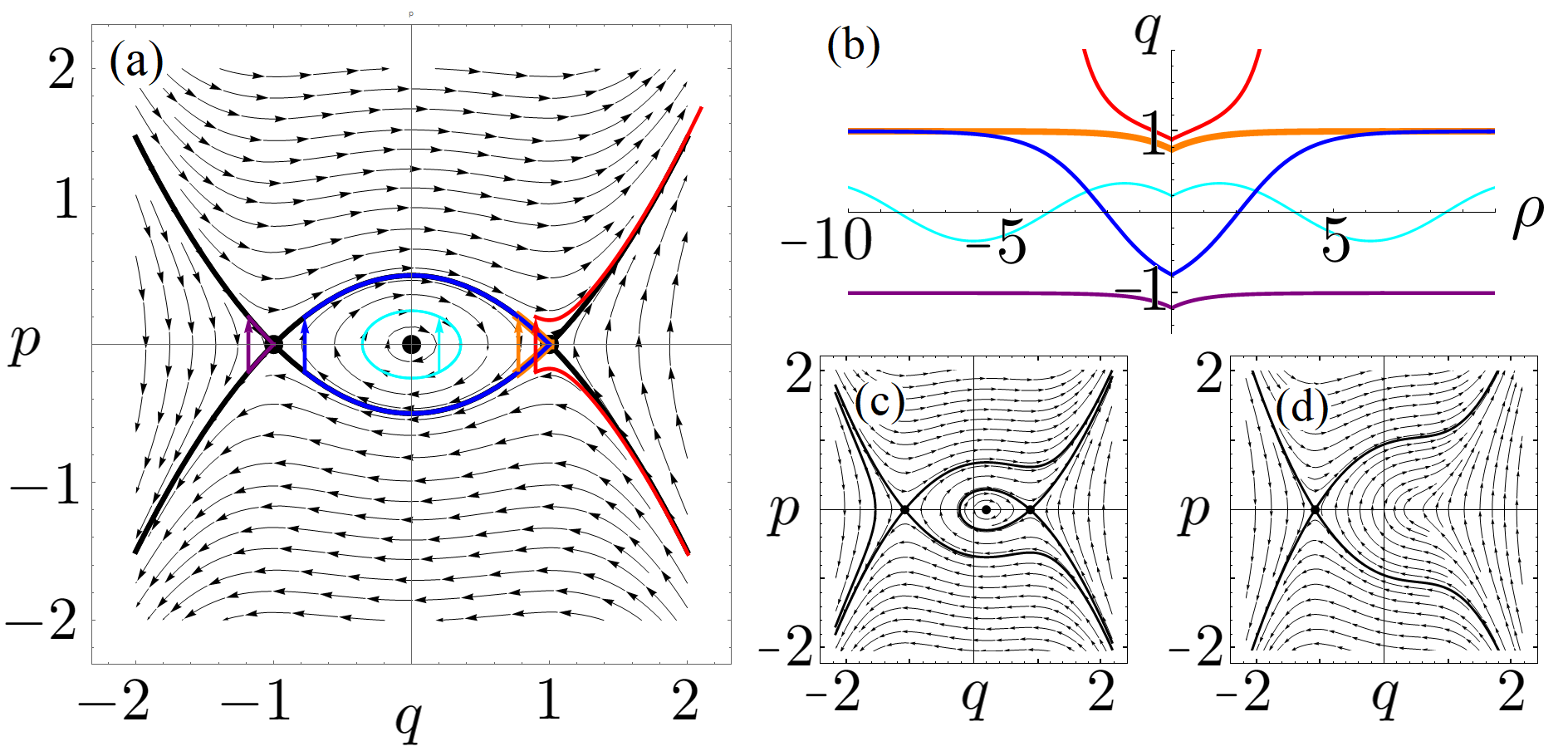}
	\caption{(a) The 1D source-free case contains three fixed points (black dots). The island boundaries and separatrices (thick black) are given by $p^2=\frac{1}{4}(q^2-1)^2$. The $\delta$-function source gives solutions a kick $\Delta p=A$ at $\rho=0$. 
	(b) Except for the three special cases $q_+^+$ (orange), $q_-^+$ (blue), and $q_-^-$ (purple), kicks either land inside the island resulting in oscillatory solutions (cyan), or land outside the island causing divergent solutions (red). 
	(c) A constant source $S_0<1/3\sqrt{3}$ deforms the phase-space island and maintains three fixed points. (d) A constant $S_0>1/3\sqrt{3}$ causes two fixed points to annihilate and destroys the phase-space island. 
	The island boundaries and separatrices (thick black) that pass through the fixed point $(Q,0)$, which satisfies $\frac{1}{2}Q(Q^2-1)+S_0=0$, are given by $p^2=\frac{1}{4}(q-Q)^2[(q+Q)^2+2(Q^2-1)]$. The solutions are in general of infinite energy.
	}
	\label{fig:portrait}
\end{figure}

In this regime, the source term may be approximated as a delta function, which gives the solution an instantaneous kick at the origin. Away from $\rho=0$, the 1D equation $\ddot{q}=q(q^2-1)/2$ can be integrated to give $d_\rho(p^2-\frac{1}{4}q^4+\frac{1}{2} q^2)=0$. For FEPs, $q\rightarrow\pm1$ and $p\rightarrow 0$ when $\rho\rightarrow\infty$, so the integration constant is $1/4$. Then, the FEPs satisfy $p^2=\frac{1}{4}(q^2-1)^2$ in the $q$-$p$ phase space (Fig.~\ref{fig:portrait}a, thick black),
which can be further integrated to find $q(\rho)$. The four possible solutions are $q=\pm\tanh\frac{1}{2}(\rho-\rho_0)$ and $q=\pm\coth\frac{1}{2}(\rho-\rho_0)$, where $\rho_0$ is some constant. 
Connecting solutions for $\rho>0$ and $\rho<0$ with the conditions that $q(0^+)=q(0^-)$ and $p(0^+)-p(0^-)=A>0$, three solutions can be constructed. The solution with $q<-1$ is (Fig.~\ref{fig:portrait}, purple)
\begin{equation}
\label{eq:q--}
q_-^-=\frac{q_c-\tanh |\rho|/2}{1-q_c \tanh|\rho|/2}, \quad q_c=-\sqrt{1+A},
\end{equation}
whose asymptotic value is $-1$ when $\rho\rightarrow\infty$. Notice that the $q_-^-$ solution exist for all $A\ge0$. The remaining two solutions satisfy $|q|<1$ and are given by
\begin{equation}
\label{eq:q-+}
q_\pm^+=\frac{q_c+\tanh |\rho|/2}{1+q_c \tanh|\rho|/2}, \quad q_c=\pm\sqrt{1-A},
\end{equation}
whose asymptotic values are both $+1$ when $\rho\rightarrow\infty$. Notice that the $q_\pm^+$ solutions exist only when $A\le1$. 
Based on the behaviors of $q_c$, 
the boundary between the 1110 and 0010 phases is given by Eq.~(\ref{eq:A_1110^0010}). 
The $q_+^+$ solution (Fig.~\ref{fig:portrait}, orange) is perturbative when $A\rightarrow0$. In contrast, the $q_-^+$ solution (Fig.~\ref{fig:portrait}, blue), which hops from $q\sim-1$ to $q\sim1$, is nonperturbative. The hopping occurs at the light horizon whose radius can be solved from $q_-^+(\rho_c)=0$, and the solution gives Eq.~(\ref{eq:rc_1DRs}). 
The light horizon thickness is $\Delta\rho\sim2$, which is on the order of the Compton wavelength. When $A$ is slightly below $1$, the order parameter $\rho_c\sim2\sqrt{1-A}$ shows a typical phase transition behavior and the critical exponent is $1/2$. When $A$ further decreases towards $0$, the light horizon radius diverges as $\rho_c\sim\ln(4/A)$, which manifests the nonperturbative nature of $q_-^+$.

The above results have three notable features. 
First, there exist two solutions with the same boundary condition $q(\infty)=1$ while only one solution with $q(\infty)= -1$ for $A>0$. In other words, the sign of the source term dictates the behaviors of FEPs, which are not uniquely determined by the boundary condition. 
Second, the potential due to a localized source is not necessarily short ranged as the Yukawa potential $\Delta q\sim \exp(-|\rho|)$. For the nonperturbative $q_-^+$, the influence of the source, which causes FEPs to deviate from the VEVs, is barely felt until possibly a large distance $\rho_c$ away. Nevertheless, notice that the depletion of $q^2$ only occurs within a thin layer whose thickness is on the order of the Compton wavelength. 
Third, the $q_\pm^+$ solutions cease to exist when the source strength is too large. In other words, for a strong source, there is no FEP that has the asymptotic value $q(\infty)=1$. In fact, when $A>1$, any potential with $q>-1$ at some $\rho$ will diverge at a finite distance.

The three FEPs have different total energy: The ground state is $q_-^-$, the first excited state is $q_+^+$, and the top state is $q_-^+$. 
To evaluate the total energy, substitute the field configurations [Eqs.~(\ref{eq:q--}) and (\ref{eq:q-+})] into the normalized energy density [Eq.~(\ref{eq:energy_density})], which can be simplified using the fact that $p^2=\frac{1}{4}(q^2-1)^2$. Then, integrating over $\rho$, which picks up half the contribution from the $\delta$-function source, the normalized energy is
\begin{equation}
\label{eq:energy_1D}
H_-^-=\frac{1}{3}(1+q_c^3),\quad H_\pm^+=\frac{1}{3}(1-q_c^3),
\end{equation}
where $q_c$ is the critical initial condition of $q_-^-$ and $q_\pm^+$, respectively. Notice that $H_-^-\le0\le H_+^+ \le H_-^+$. Moreover, $H_-^-$ and $H_-^+$ are monotonously decreasing functions of A, whereas $H_+^+$ is monotonously increasing. When $A\rightarrow0$, the lowest two states become degenerate with $H_-^-=H_+^+=0$, whereas the excited state $H_-^+=2/3$ reaches the highest energy. On the other hand, when $A\rightarrow1$ approaches the 1110-0010 phase boundary, the two excited states become degenerate with $H_-^-=H_+^+=1/3$. Beyond the phase boundary, the $q_\pm^+$ solutions no longer exist, while $H_-^-$ keeps on decreasing.

\subsection{Asymptotic solutions for $R\gg 1$ \label{sec:1D:constant}}
In this regime, the source is much larger than the Compton wavelength. Therefore, the instantaneous fixed points, which satisfy $\frac{1}{2}q(q^2-1)+S=0$, are being dragged adiabatically as the source slowly ramps off. 
Notice that there are three real roots when $0\le S<1/3\sqrt{3}$, and only one real root when $S>1/3\sqrt{3}$. 
Therefore, the 0010 phase boundary is given by Eq.~(\ref{eq:A_1110^2010}). 
Below the phase boundary, the adiabatic solutions can be written as 
\begin{equation}
\label{eq:adiabatic}
q_a=\frac{2}{\sqrt{3}}\cos\frac{\psi+2\pi n}{3}, 
\end{equation}
where $\psi=\arccos(-3\sqrt{3}S)$ is a function of $\rho$ and $n$ is an integer. When $\rho\rightarrow\infty$, $S\rightarrow 0$, so $q_a(n=0)\rightarrow 1$, $q_a(n=1)\rightarrow -1$, and $q_a(n=2)\rightarrow 0$. Therefore, only the $n=0$ and $n=1$ solutions, which will be referred to as $q_a^+$ and $q_a^-$, correspond to FEPs. 
To see under what conditions is Eq.~(\ref{eq:adiabatic}) the dominant balance of Eq.~(\ref{eq:EOM}), take $\rho$ derivative on both sides of the cubic equation, which gives $\dot{q}_a=2\dot{S}/(1-3q_a^2)$ and $\ddot{q}_a=2\ddot{S}/(1-3q_a^2)+24q_a\dot{S}^2/(1-3q_a^2)^3$. For the Gaussian source, $\dot{S}=-2\rho S/R^2$ and $\ddot{S}=(2S/R^2)(2\rho^2/R^2-1)$, so the left hand side (LHS) of Eq.~(\ref{eq:EOM}) becomes $\ddot{q}_a+(D-1)\dot{q}_a/\rho=\{4S/[R^2(1-3q_a^2)]\}\{(2\rho^2/R^2)[1+12 q_aS/(1-3q_a^2)^2]-D\}$. 
We see LHS $\ll S$ when $R^2(3q_a^2-1)\gg1$, if the term in the second braces is $O(1)$. In other words, for sufficiently large $R$ and away from $q^2=1/3$, the adiabatic solutions asymptotically solve Eq.~(\ref{eq:EOM}) within the source region. On the other hand, outside the source region, $\Delta q$ approaches the Yukawa potential, which decays exponentially on the scale of the Compton wavelength.

The adiabatic solutions are valid in all spatial dimensions, and their total energy is easy to estimate in the perturbative regime. Notice that when both $q_a^\pm$ exist, $S_0\le1/3\sqrt{3}\approx0.2$ is small. Therefore, we can treat $S$ as a perturbation. Then, to lowest order, the adiabatic solutions $q_a^\pm\simeq\pm1-S$, and the normalized energy density $\mathcal{E}\simeq\pm S$. Integrating the Gaussian source in $D$-dimensional space, the total energy is
\begin{equation}
\label{eq:energy_adiabatic}
H_a^\pm\simeq \pm \frac{A}{2\pi^{D/2}}\Gamma\Big(\frac{D}{2}\Big),
\end{equation}
where $\Gamma$ is the gamma function. In the perturbative regime, the two adiabatic FEPs have approximately equal and opposite energy, which is independent of $R$, and the energy split increases linearly with $A$ .

In addition to the adiabatic solutions, there also exists a nonperturbative solution that hops between the VEVs. 
Similar to Sec.~\ref{sec:1D:delta}, when $A>0$, the hopping solution can only jump from $-1$ to $+1$, but not vice versa. When hopping occurs on a scale that is much smaller than $R$, the source term may be regarded as a constant. For $S_0<1/3\sqrt{3}$, let $1/\sqrt{3}<q_\infty<1$ be the largest fixed point. Then, integrating the 1D equation $\ddot{q}=q(q^2-1)/2+S_0$, the FEP with $q(\infty)=q_\infty$ satisfies $p^2=\frac{1}{4}(q-q_\infty)^2(q-q_b)(q-q_c)$, where $q_b=-q_\infty-f$, $q_c=-q_\infty+f$, and $f=\sqrt{2(1-q_\infty^2)}$. Further integrating this equation for $q_c<q<q_\infty$ gives the hopping solution
\begin{equation}
\label{eq:1D_hopping}
q_{h}=\frac{(2+\theta)q_c+2q_\infty\sinh^2k\rho}{\theta+2\cosh^2k\rho},
\end{equation} 
where $\theta=(q_\infty-q_c)/(q_\infty+q_c)$ and $k=\sqrt{(3q_\infty^2-1)/8}$. We see $q_h(\rho)$ varies on the scale $\Delta\rho\sim1/k$, which goes to infinity when $q_\infty\rightarrow1/\sqrt{3}$. The hopping solution monotonously increases from the initial condition $q(0)=q_c$, which lies between the two smaller fixed points, to the largest fixed point $q_\infty$. 
After reaching $q_\infty$, the solution then tracks the adiabatic solution $q_a^+$ as the source slowly ramps off. 
The above approximations is the dominant balance of Eq.~(\ref{eq:EOM}) when $kR\gg1$. In other words, as long as the adiabatic solutions are good approximations, $q_h$ is also a good asymptotic solution.

The hopping solution exhibits a light horizon when $q_c<0$. This occurs when $q_\infty>\sqrt{2/3}$, which corresponds to $S_0<1/3\sqrt{6}$. Therefore, the 1110-2010 phase boundary is given by Eq.~(\ref{eq:A_1110^2010}). The light horizon radius can be solved from $q_h(\rho_c)=0$, and the solution $\rho_c$ gives Eq.~(\ref{eq:rc_1DRl}). 
To see the asymptotic behavior of $\rho_c$ when $S_0\rightarrow 0$, notice that $q_\infty\sim1-S_0$, $f\sim2\sqrt{S_0}$, $q_c\sim-1$, $\theta\sim1/\sqrt{S_0}$, and $k\sim\frac{1}{2}$. Since arcsinh$\,x\sim\ln(2x)$ when $x\rightarrow+\infty$, we see $\rho_c\sim \ln(2/\sqrt{S_0})$ diverges logarithmically when $S_0\rightarrow 0$.  
In the opposite limit, when $S_0$ increases towards the 1110-2010 phase boundary $S_c=1/3\sqrt{6}$, we can write $q_\infty\sim\sqrt{2/3}+\varepsilon$ where $\varepsilon=2(S_c-S_0)\rightarrow 0$. Then, $f\sim\sqrt{2/3}-2\varepsilon$, $q_c\sim-3\varepsilon$, $\theta\sim1$, and $k\sim1/2\sqrt{2}$. Since arcsinh$\,x\sim x$ when $x\rightarrow0$, we see $\rho_c\sim 6^{3/4}\sqrt{3\varepsilon}=6^{5/4}\sqrt{S_c-S_0}$.

The energy of the hopping solution, which is always larger than the energy of adiabatic solutions, has a simple approximation when $S_0\rightarrow0$.
To lowest order, $q_h\simeq(\sinh^2\rho/2-V)/(\cosh^2\rho/2+V)$, where $V=1/2\sqrt{S_0}$. When $V\gg1$, the kinetic energy $\frac{1}{2}p^2$ roughly equals to the self potential energy $\frac{1}{8}(q^2-1)^2$, which dominates the external potential energy $Sq$. Then, the energy density $\mathcal{E}_h\simeq V^2(\sinh^2\rho/2+\cosh^2\rho/2)^2/(\cosh^2\rho/2+V)^4$. 
The integral $H_h=\int_0^\infty d\rho\mathcal{E}_h$ can be carried out exactly. However, since the approximation is only valid when $V\gg1$, it is sufficient to keep the leading terms 
\begin{equation}
\label{eq:energy_hopping1D}
H_h\simeq\frac{2}{3}\Big(1-\frac{1}{V}+\frac{3\text{arcsinh}\sqrt{V}}{2V^2}+\dots\Big).
\end{equation}
We can write $H_h=2g/3$, where the factor $g(V)\sim 1$. In other words, $H_h$ only has a weak dependence on $S_0=1/4V^2$. Check that in the limit $S_0\rightarrow0$, $H_h\rightarrow2/3$ coincides with the case of a $\delta$-function source $H_-^+$ [Eq.~(\ref{eq:energy_1D})]. This is expected because when the source vanishes, the energy of the nonperturbative FEP is agnostics of the source size.

While exhibiting some notable differences, solutions for $R\gg1$ have many features in common with $R\ll1$. In both cases, when the source is weak, there exists a single FEP with $q(\infty)= -1$, and two FEPs with $q(\infty)=1$, one of which is perturbative and the other is nonperturbative. 
The nonperturbative FEP is the highest-energy state and hops from the lower VEV to the upper within a layer of thickness $\Delta\rho$. When the critical initial condition $q_c<0$, the hopping solution exhibits a light horizon whose radius $\rho_c\sim\ln(1/A)$ and thickness $\Delta\rho\sim 1$ when $A\sim0$. 
As $A$ increases, $q_c$ for the hopping solution becomes positive when $R\gg1$, giving rise to a 2010 phase where the light horizon no longer exists. Nevertheless, in the 2010 phase, the scale length of the nonperturbative FEP is $\Delta\rho\rightarrow R$ when $A\rightarrow A_{2010}^{0010}$.
We see in both the 1110 and the 2010 phases, the nonperturbative FEP varies on scales that are much larger than the Compton wavelength. 
Finally, when $A$ further increases, the two FEPs with $q(\infty)=1$ extinct when $A$ exceeds the phase boundary.

\subsection{Numerical results \label{sec:1D:numeric}}
In the intermediate regime $R\sim1$, all terms of Eq.~(\ref{eq:EOM}) are comparable, and the nonlinear equation is solved numerically to mapped out the phase diagram. Notice that in 1D, $\rho$ is defined also for $\rho<0$. Therefore, to capture a more complete picture, we can relax the constraint that 
the initial condition $p(0)$ is zero. 
Then, regarding Eq.~(\ref{eq:EOM}) as a classical dynamical system, its behavior can be analyzed in the $q$-$p$ phase space.

The $q$-$p$ phase portraits are well defined for a $\delta$-function source and a constant source, for which Eq.~(\ref{eq:EOM}) has no explicit $\rho$ dependence.
For the $\delta$-function source (Fig.~\ref{fig:portrait}a), the phase portrait is the same as the source-free case, except for a $\Delta p=A>0$ kick at $\rho=0$. 
The two FEPs with $q(\infty)=1$ are the results of the two ways of kicking the solution from the lower island boundary to the upper (orange and blue). 
Since the $p$ width of the island is $1$, we see $A=1$ is the critical value beyond which no kick can both start and end on the island boundaries.
On the other hand, the FEP with $q(\infty)=-1$ always exists (purple), which is obtained by kicking the solution from the lower left separatrix to the upper. In general, the solutions are either oscillatory (cyan), which occurs inside the island, or divergent (red), which occurs outside the island. 
In the opposite limit $R\rightarrow\infty$, 
for a constant source with $S_0<1/3\sqrt{3}$, there exist three fixed points and a phase-space island (Fig.~\ref{fig:portrait}c). The perturbative FEPs are the adiabatic solutions that track the saddle points, while the nonperturbative FEP travels along the island boundary. 
In comparison, when $S_0>1/3\sqrt{3}$, only one saddle point remains (Fig.~\ref{fig:portrait}d), which gives rise to the the only FEP with $q(\infty)=-1$.
These phase-space pictures give intuitive explanations for results in Sec.~\ref{sec:1D:delta} and \ref{sec:1D:constant}.

\begin{figure}[t]
	\centering
	\includegraphics[width=0.48\textwidth]{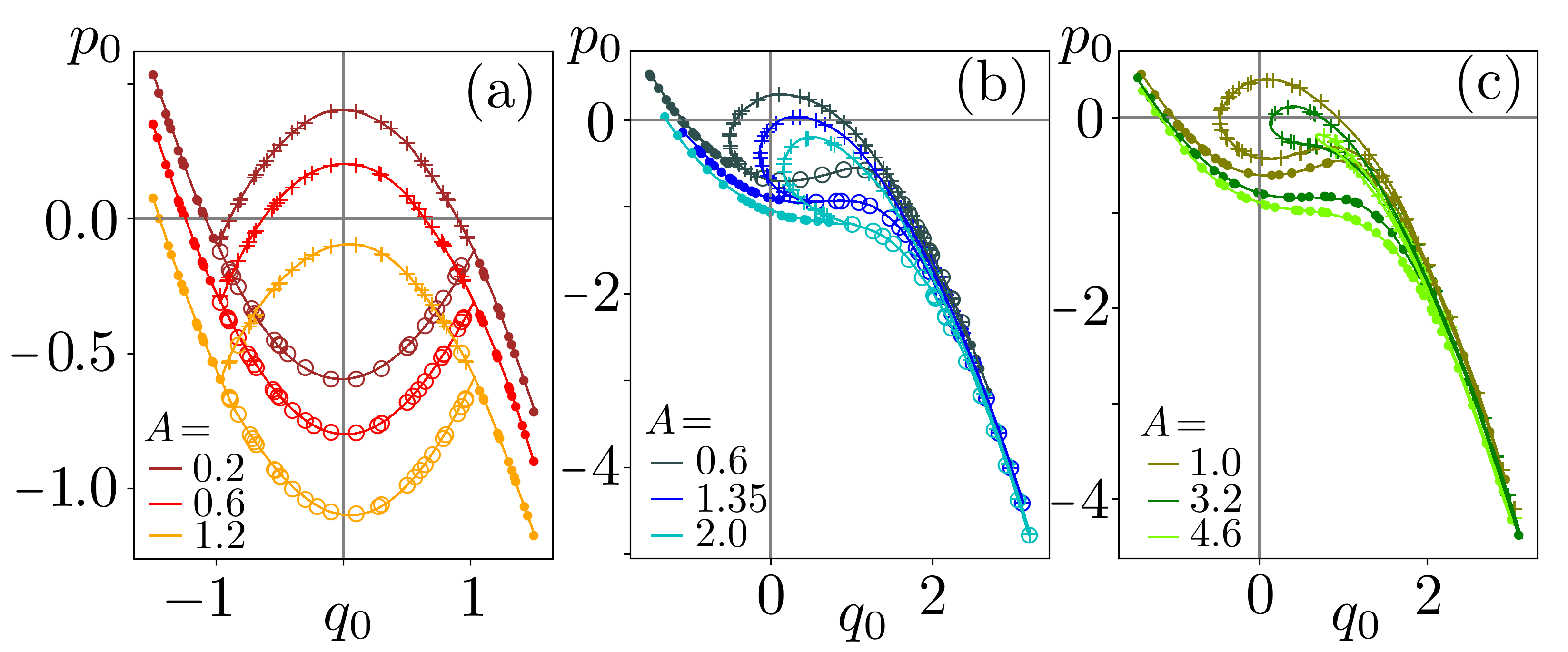}
	\caption{(a) Numerically obtained $q_0$-$p_0$ phase portraits for small $R=0.01$, which are closely related to the $\delta$-function-source case, are displaced downwards to compensate for the $\Delta p=A/2$ kick when $\rho$ increases from $0$. (b) At intermediate $R\approx2.37$, in addition to displacements, the left separatrix gradually peels off the island, whose right side slims into the right separatrix. (c) At large $R=100$, the phase portraits closely resembles the constant-source case. For larger $A$, the left separatrix detaches and the island shrinks.  
	The separatrices are marked by ``$\bullet$", which divide the island boundary into two segments. Outside the segment marked by ``+", solutions diverges to $+\infty$, while outside the boundary marked by ``$\circ$", solutions diverges to $-\infty$; Within the island, solutions are oscillatory. FEPs are solutions whose initial conditions are on island boundaries or separatrices.
	}
	\label{fig:Island}
\end{figure}

Based on the pictures for the special cases $R=0$ and $R=\infty$, it is now easy to understand the problem at finite $R$. In this case, due to the explicit $\rho$ dependence, the phase portrait is not well defined. Nevertheless, by scanning the initial conditions in the $q_0$-$p_0$ space, features that are analogous to phase-space islands and separatrices can be identified (Fig.~\ref{fig:Island}). For $(q_0,p_0)$ within the island, solutions are oscillatory when $\rho\gg R$. On the other hand, beyond the island boundaries, solutions are divergent. At finite $\rho$, $q\rightarrow+\infty$ outside the island boundary marked by ``+", while $q\rightarrow-\infty$ outside the boundary marked by ``$\circ$". The island boundary is divided into the ``+" and ``$\circ$" segments by two separatrices marked by ``$\bullet$". For $(q_0,p_0)$ above the separatrix, $q\rightarrow+\infty$, while below the separatrix, $q\rightarrow-\infty$.
When $R=0.01$ (Fig.~\ref{fig:Island}a), the portraits are similar to the $\delta$-function-source case. 
For a larger $A$, the island is displaced further downward, so that it can be restored to the source-free case by the $\Delta p=A/2$ kick when $\rho$ increases from 0.
Likewise, when $R=100$ (Fig.~\ref{fig:Island}c), the portraits closely resemble the constant-source case. 
As $A$ increases, the island shrinks and eventually vanishes. Notice that when $R\gg1$, solutions with $(q_0,p_0)$ immediately outside the island always diverges to $+\infty$. 
Finally, when $R\approx 2.37$ (Fig.~\ref{fig:Island}b), we see an intermediate behavior: As $A$ increases, the island is shifted downwards. At the same time, the left separatrix gradually peels off the island, while the right side of the island narrows and merges into the right separatrix.

\begin{figure}[t]
	\centering
	\includegraphics[width=0.48\textwidth]{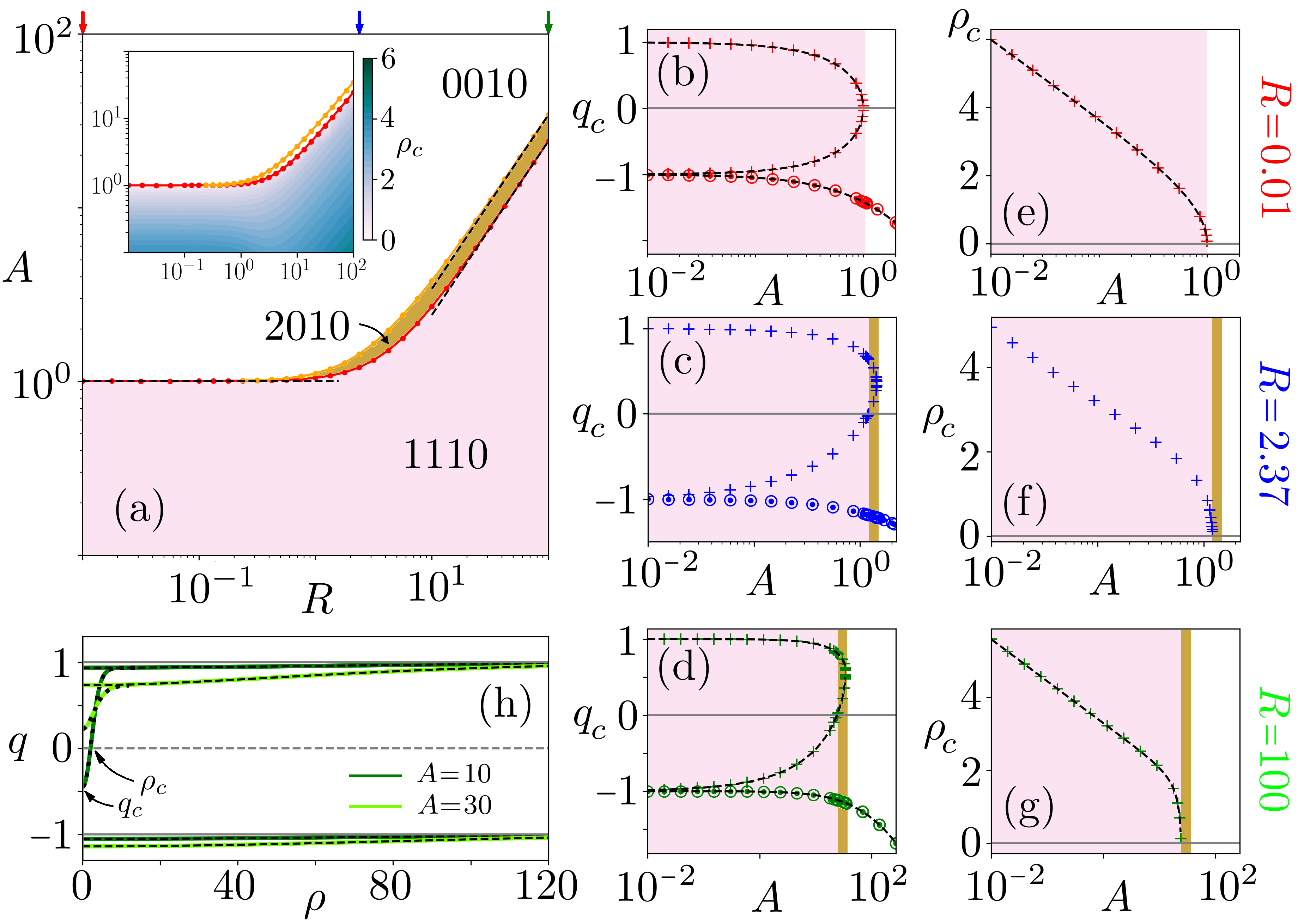}
	\caption{(a) For the 1D problem, the $R$-$A$ phase diagram is constituted of three phases. In the 1110 phase (pink), three FEPs exist, one of which is nonperturbative and exhibits a light horizon, whose radius $\rho_c$ is shown in the inset. In the 2010 phase (brown), the light horizon disappears; and in the 0010 phase (white), only the FEP with $q(\infty)=-1$ (``$\odot$" marker) remains. 
	As illustrated by the three examples at $R=0.01$ (b, e), $R\approx2.37$ (c, f), and $R=100$ (d, g), the critical initial condition $q_c$ and the light horizon radius $\rho_c$ are two order parameters:
	The two FEPs with $q(\infty)=+1$ (``+" marker) annihilate when $A$ reaches the 0010 phase;
	$\rho_c$ terminates when $A$ enters the 2010 phase.
	(h) At $R=100$, example solutions when $A$ is in the 1110 (dark green) and 2010 (light green) phases match asymptotic expressions when the FEP is the adiabatic [Eq.~(\ref{eq:adiabatic}), dashed] and the hopping [Eq.~(\ref{eq:1D_hopping}), dotted] solutions. 
	Black lines are asymptotic results, and colored markers are numerical data. 
	}
	\label{fig:1D}
\end{figure}

Having understood the 1D problem as a classical dynamical system, we can return to the analysis of spherically symetric potentials, for which $p(0)=0$. Then, it is sufficient to focus on the $q_0$ axis, where the islands and separatrices cross at $q_c$. Solutions with the initial condition $(q_0, p_0)=(q_c,0)$ approach $q(\infty)=\pm1$, and yield FEPs. 
From the $q_0$-$p_0$ phase portraits, it is easy to see that three FEPs exist for small A, two of which come from the upper island boundary while the third comes from the left separatrix. On the other hand, when $A$ increases above a threshold, only the separatrix FEP remains. 
The threshold, namely, the 0010 phase boundary (Fig.~\ref{fig:1D}a, orange) is mapped out by bisecting $A$ at a given $R$, and numerical results match the asymptotic results (dashed black) given by Eqs.~(\ref{eq:A_1110^0010}) and (\ref{eq:A_1110^2010}). 
To illustrate the discontinuous nature of the phase transition, three examples of $q_c(A)$ are shown in Figs.~\ref{fig:1D}b-\ref{fig:1D}d. In these figures, the island boundaries are marked by ``+", indicating $q(\infty)=+1$, and the separatrices are marked by ``$\odot$", indicating $q(\infty)=-1$. The dashed black lines are asymptotic results, which match numerical results for $R\ll1$ and $R\gg1$.
Regarding $q_c$ as an order parameter, we see the two upper FEPs annihilate when $A$ reaches the 0010 phase.

In addition to discrete jumps of the number of FEPs, softer phase transitions can also be identified using the light horizon radius $\rho_c$ as another order parameter (Fig.~\ref{fig:1D}a, inset). Since $q(\rho)$ is continuous and monotonous, 
it is easy to see that $\rho_c$ exists only for the ``+" solutions when $q_c\le 0$. The condition $q_c=0$ determines the 1110-2010 phase boundary (Fig.~\ref{fig:1D}a, red), whose asymptotic expression is given by Eq.~(\ref{eq:A_1110^2010}) when $R\gg1$ (dashed black).
To illustrate the 1110-2010 phase transition, examples of $\rho_c(A)$ are shown in Figs.~\ref{fig:1D}e-\ref{fig:1D}g, where the dashed black lines are asymptotic results [Eqs.~(\ref{eq:rc_1DRs}) and (\ref{eq:rc_1DRl})].
When $A$ increases towards the phase boundary $A_c$, instead of gradually falling off, the light horizon radius terminates sharply as $\rho_c\sim\sqrt{A_c-A}$. 
In the opposite limit, $\rho_c\sim\ln(1/A)$ diverges logarithmically when $A\rightarrow 0$, which is clearly a nonperturbative behavior. 
Across the light horizon, the FEP passes $q=0$ within a layer of thickness $\Delta\rho\sim O(1)$, and an example solution is shown in Fig.~\ref{fig:1D}h (dark green) when $R=100$ and $A=10$, where the dotted black line is the asymptotic result [Eq.~(\ref{eq:1D_hopping})]. 
The thickness of the hopping layer, where $q(\rho)$ changes rapidly, increases with $A$. After $A$ reaches the 2010 phase, the light horizon disappears, but the nonperturbative FEP continues to exist, for which an example is shown in Fig.~\ref{fig:1D}h (light green) when $A=30$. The thickness $\Delta\rho\sim O(R)$ when $A$ increases towards the 2010-0010 phase boundary, beyond which the two FEPs with $q(\infty)=+1$ annihilate.
We see whenever the nonperturvative FEP exists, it varies on scales that are much larger than the Compton wavelength.

\section{Potentials in higher dimensions \label{sec:DD}}
In higher dimensions, although nontrivial vacuum field configuration does not exist according to Derrick's theorem, nonperturbative FEPs are possible in the presence of a source term, which allow tunneling between the VEVs. In contrast to the 1D case where there is at most one nonperturbative FEP, up to three nonperturbative FEPs can exist in higher dimensions. Moreover, in higher dimensions, there always exist at least one FEP with $q(\infty)=+1$, unlike in the 1D case where such a potential ceases to exist for sufficiently large $A$. 
The differences originate from the singular term $p/\rho$, which substantially changes the $R$-$A$ phase diagram. 
Nevertheless, some essential features are preserved: For given boundary conditions, the FEPs are not always unique, and nonperturbative FEPs, whenever exist, span on scales that are much larger than the Compton wavelength.

Key results that will be shown in this section are summarized here: (i) When $R\ll1$, the system transitions from the 1010 phase to the 0110 phase when $A$ increases beyond $A_{1010}^{0110}$. In both phases, there is a unique FEP with $q(\infty)=-1$ and a unique FEP with $q(\infty)=+1$.
The phase boundaries in 2D and 3D are
\begin{eqnarray}
\label{eq:Ac_Rs_2D}
A_{1010}^{0110}(D=2)&\simeq&\frac{2\pi}{\ln(2/R)-\gamma/2},\\
\label{eq:Ac_Rs_3D}
A_{1010}^{0110}(D=3)&\simeq&\frac{2\pi R}{1/\sqrt{\pi}-R/2},
\end{eqnarray}
where $\gamma$ is the Euler–Mascheroni constant. Notice that $A_{1010}^{0110}\rightarrow0$ when $R\rightarrow 0$, which means that a $\delta$-function source is always in the 0110 phase. In this phase, one FEP exhibits a light horizon, whose radius $\rho_c$ is on the order of the Compton wavelength.
(ii) When $R\gg1$, the $D$-dimensional system transitions though 1010, 1210, 2110, and 0110 phases as $A$ increases, where the phase boundaries are given by
\begin{eqnarray}
\label{eq:A_1010^1210}
A_{1010}^{1210} &\simeq& \frac{D-1}{3}\sqrt{2e\pi^D} R^{D-1},\\
\label{eq:A_2110^0110}
A_{1210}^{2110}&\lesssim& A_{2110}^{0110}\simeq \frac{\pi^{D/2}}{3\sqrt{3}}R^D,
\end{eqnarray}
where $e$ is the base of the natural logarithm. Notice that $A_{1010}^{1210}\propto R^{D-1}$ while $A_{1210}^{2110}\propto R^D$. Therefore, as $R$ increases, the onset of the 1210 phase occurs at lower density $S_0$, and the 1210 phase becomes wider.
In the 1210 phase, two FEPs hop from $q<0$ to $q>0$ at different radii $\rho_\pm$, whose lowest-order approximations are 
\begin{eqnarray}
\label{eq:rho+}
\rho_+&\simeq& R\sqrt{\ln\frac{3R S_0}{D-1}},\\
\label{eq:rho-}
\rho_-&\simeq& \frac{D-1}{3S_0},
\end{eqnarray}
when $\rho_\pm$ is not much larger than $R$. 
At the 1010-1210 phase boundary, the exact values of $\rho_\pm$ equal. As $S_0$ increases from the boundary, $\rho_+$ grows logarithmically, while $\rho_-$ drops to zero at the 1210-2110 phase boundary, near which Eq.~(\ref{eq:rho-}) is no longer a good approximation.
The two nonperturbative FEPs have different energy $H_+<H_-$, which means that a larger light horizon is always energetically more favorable. Near the 1010-1210 boundary, $H_a^-<H_a^+<H_+<H_-$, which means that the adiabatic FEP $q_a^+$ is the first excited state. However, for slightly larger $A$ values, the energy levels cross and $H_a^-<H_+<H_a^+<H_-$, which means the nonperturbative FEP $q_+$ becomes the first excited state. 
(iii) Numerical results are summarized in Fig.~\ref{fig:2D} and Fig.~\ref{fig:3D}. When $R\sim1$, there exist an additional 0310 phase, wherein three FEPs exhibit light horizons. In total, there are five phases, and in every phase there are at least two FEPs, one with $q(\infty)=-1$ and the other with $q(\infty)=+1$. Except for the 1010 phase where $A$ is small, all other phases have at least one FEP with a light horizon. Contrary to what happens in 1D, light horizons emerge when the source strength is above a critical value in higher dimensions.

\subsection{Asymptotic solutions for $R\ll1$ \label{sec:DD:delta}}
Unlike the 1D case, a $\delta$-function source does not capture the full picture when $R\rightarrow 0$ in higher dimensions. This is because the term $(D-1)p/\rho$ diverges when $\rho\rightarrow 0$, which competes with the divergence of the source term when $R\rightarrow 0$. 
Asymptotic solutions for $R\ll1$ may be divided into several regions, and in each region, the dominant balance of Eq.~(\ref{eq:EOM}) is given by a different set of terms. In particular, near the phase boundary $A\sim A_{1010}^{0110}$, the solutions can be divided into inner and outer regions. In the inner region, the nonlinear term can be ignored; while in the outer region, the source term is subdominant and the nonlinear term can be linearized. 
By matching the inner and outer solutions in their common region of validity, global asymptotic solutions can be constructed. 
The exact behaviors in 2D and 3D are different, which will be discuss below separately.

\subsubsection{2D solutions near phase transition \label{sec:DD:delta:2D}}  
In the inner region, the source term diverges when $R\rightarrow 0$ for a fixed $A$. However, $q\sim O(1)$ remains finite near the 1010-0110 phase boundary, which means that the nonlinear term is subdominant. Then, the dominant balance of Eq.~(\ref{eq:EOM}) is given by the linear equation $\ddot{q}+\dot{q}/\rho\simeq S$. 
The homogeneous problem can be easily integrated to give $p\propto 1/\rho$. For the inhomogeneous problem, suppose $p(\rho)=c(\rho)/\rho$, then $\dot{c}=\rho S$. This equation can be integrated for the Gaussian source to give $c=c_0+(A/2\pi)[1- \exp(-\rho^2/R^2)]$. The integration constant $c_0=0$, because of the initial condition $p(0)=0$ for spherically symmetric potentials.
Then, for the dominant balance to hold near the 1010-0110 phase boundary, the condition $p/\rho\gg1$ is satisfied if $\rho\ll\sqrt{A/2\pi}$.
Within this region where the above dominant balance is valid, further integrating $p=\dot{q}$ with $q(0)=q_0$ gives the asymptotic solution in the inner region
\begin{equation}
\label{eq:q_in_2D}
q_{\text{in}}\simeq q_0+\frac{A}{4\pi}\Big[\gamma+\ln\frac{\rho^2}{R^2}+E_1\Big(\frac{\rho^2}{R^2}\Big)\Big],
\end{equation}
where $\gamma$ is the Euler–Mascheroni constant and $E_1(z)=\int_z^\infty dt \exp(-t)/t$ is an exponential integral. 
Check that in the limit $z\rightarrow0$, $E_1(z)\simeq-\gamma-\ln z + z+\dots$, so $q_{\text{in}}\rightarrow q_0$ as demanded. 
In the opposite limit $z\rightarrow\infty$, $E_1(z)\simeq\exp(-z)/z+\dots$ is subdominant, so $q_{\text{in}}\simeq (A/2\pi)\ln\rho + q_0+(A/4\pi)(\gamma-2\ln R) +\dots$. We see that $q_{\text{in}}$ depends on $\rho$ logarithmically when $\rho\rightarrow\infty$. This asymptotic behavior will be used to match with the outer solution.

In the outer region $\rho\gg R$, the source term $S\propto \exp(-\rho^2/R^2)$ falls off rapidly, so the equation quickly approaches the source-free problem. According to Derrick's theorem, there is no nontrivial vacuum field configurations in 2D. Therefore, the FEPs must be close to the VEVs. We can then expand $q=\pm1+y$, where $|y|\ll1$ is perturbative. To lowest order, $y$ satisfies $\ddot{y}+\dot{y}/\rho-y\simeq 0$, and the general solutions are given by the modified Bessel functions $K_0$ and $I_0$. Using the boundary condition $y\rightarrow 0$ when $\rho\rightarrow\infty$, only $K_0$ contributes, so $y\simeq B K_0(\rho)$ where $B$ is some constant. 
To determine $B$, notice that in the limit $\rho\rightarrow 0$, $K_0(\rho)\simeq-\ln\rho+(\ln2-\gamma)+\dots$. In other words, $q_{\text{out}}(\rho\rightarrow0)$ matches the $\ln\rho$ behavior of $q_{\text{in}}(\rho\rightarrow\infty)$. Asymptotic matching gives $B=-A/2\pi$, and the outer solution is the Yukawa potential
\begin{equation}
q_{\text{out}}\simeq\pm 1 - \frac{A}{2\pi} K_0(\rho).
\end{equation} 
Matching the inner and outer solutions to the next order, $A$ and $q_0$ become related. In other words, similar to the 1D case, the critical initial conditions that lead to FEPs are not arbitrary, but are given by
\begin{equation}
q_c \simeq \pm1-\frac{A}{2\pi}\Big(\ln\frac{2}{R}-\frac{\gamma}{2}\Big).
\end{equation}
Substituting $q_c$ into Eq.~(\ref{eq:q_in_2D}), 
$q_{\text{in}}\simeq\pm1+(A/2\pi)[\gamma+\ln(\rho/2)+\frac{1}{2}E_1(\rho^2/R^2)]$. 
While the ``$-$" solution is always below $-1$, the ``+" solution, which satisfies $q(\infty)=+1$, exhibits a light horizon with radius $\rho_c\sim O(1)$ whenever $q_c<0$. Hence, the 1010-0110 phase boundary is given by the condition $q_c=0$, which yields the critical source strength $A_c$ in Eq.~(\ref{eq:Ac_Rs_2D}).  
When $A\rightarrow A_c$, $\rho_c$ may be approximated as the root of $q_{\text{in}}=0$, whose limiting form is $\rho_c\rightarrow 2R\sqrt{\pi(1/A_c-1/A)}$.
Finally, as \textit{a posteriori} check, notice that $A_c$ decreases as $1/\ln (1/R)\gg R^2$ when $R\rightarrow 0$. Therefore, the region of validity for the inner solution, which is given by $\rho\ll\sqrt{A/2\pi}$, overlaps better with the region of validity for the outer solution, which is given by $\rho\gg R$. Consequently, the asymptotic expressions give better approximations when the source size is smaller.

\subsubsection{3D solutions near phase transition \label{sec:DD:delta:3D}}  
Following similar procedures, we can find the asymptotic solutions in 3D near the 1010-0110 phase boundary. In the inner region,  
the dominant balance of Eq.~(\ref{eq:EOM}) is $\ddot{q}+2\dot{q}/\rho\simeq S$. 
The homogeneous problem is solved by $p\propto 1/\rho^2$, and the inhomogeneous problem is solved by $p(\rho)=c(\rho)/\rho^2$ where $\dot{c}=\rho^2 S$. Integrating $\dot{c}$ with the initial condition $p(0)=0$, we have $c=(A/\pi^{3/2})[(\sqrt{\pi}/4)\text{erf}(\xi)-(\xi/2)\exp(-\xi^2)]$, where $\xi=\rho/R$ and $\text{erf}(\xi)=(2/\sqrt{\pi})\int_0^\xi dt \exp(-t^2)$ is the error function. 
Further integrating $p=\dot{q}$ with the initial condition $q(0)=q_0$ yields the inner solution
\begin{equation}
\label{eq:q_in_3D}
q_{\text{in}}\simeq q_0+\frac{A}{2R\pi^{3/2}}\Big[1-\frac{\sqrt{\pi}R}{2\rho} \text{erf}\Big(\frac{\rho}{R}\Big)\Big],
\end{equation}
which is valid for $\rho\ll(A/2\pi)^{1/3}$. 
When $\rho\rightarrow0$, $\text{erf}(\xi)\simeq(2/\sqrt{\pi})(\xi-\xi^3/3+\dots)$, so $q_{\text{in}}\rightarrow q_0$ as demanded. 
In the opposite limit $\rho\rightarrow\infty$, $\text{erf}(\xi)\simeq1-\exp(-\xi^2)/\sqrt{\pi}\xi+\dots$, so $q_{\text{in}}\simeq q_0+(A/2\pi^{3/2})(1/R-\sqrt{\pi}/2\rho+\dots)$, which will be used to match with the outer solution.

In the outer region, 
we can expand $q=\pm1+y$ where $|y|\ll1$. To lowest order, $y$ satisfies $\ddot{y}+2\dot{y}/\rho-y\simeq 0$, and the general solutions are given by the modified spherical Bessel functions $k_0$ and $i_0$. Using the boundary condition $y(\infty)=0$, only $k_0$ contributes so $y\simeq B k_0(\rho)$, where $B$ is determined by asymptotically matching. Using the exact special case $k_0(\rho)=\pi\exp(-\rho)/2\rho$, the solution in the outer region $\rho\gg R$ is given by the Yukawa potential
\begin{equation}
\label{eq:q_out_3D}
q_{\text{out}}\simeq\pm 1 - \frac{A}{4\pi\rho} e^{-\rho}.
\end{equation} 
To the next order, matching the inner and outer solutions gives the critical initial conditions for the two FEPs
\begin{equation}
\label{eq:qc_3D}
q_c \simeq \pm1+\frac{A}{2\pi}\Big(\frac{1}{2}-\frac{1}{\sqrt{\pi}R}\Big),
\end{equation}
which can be substituted into Eq.~(\ref{eq:q_in_3D}) to give $q_{\text{in}}\simeq\pm1+(A/4\pi)[1-\frac{1}{\rho}\text{erf}(\rho/R)]$. 
The ``$-$" solution, which satisfies $q(\infty)=-1$, is always below $-1$. In contrast, the ``+" solution, which satisfies $q(\infty)=1$, crosses zero at $\rho_c\sim O(1)$ if $q_c<0$. Hence, the 1010-0110 phase boundary is given by the condition $q_c=0$, which gives Eq.~(\ref{eq:Ac_Rs_3D}). 
Near the phase boundary, $\rho_c$ is well approximated by the root of $q_\text{in}=0$, 
whose limiting behavior is $\rho_c\rightarrow R \sqrt{(6\pi^{3/2}R)(1/A_c-1/A)} \simeq R\sqrt{3(1-A_{c}/A)}$.
Similar to the 2D case, the asymptotic expressions give better approximations for smaller $R$.

\subsection{Asymptotic solutions for $R\gg1$ \label{sec:DD:constant}}
Similar to the 1D case, there exist two adiabatic solutions [Eq.~(\ref{eq:adiabatic})], 
which track the saddle points, as well as hopping solutions, which tunnel from $q\sim-1$ to $q\sim+1$. However, due to the singular term $p/\rho$, 1D-like hopping can no longer occur near $\rho\sim0$. This is because for 1D-like hopping, $q$, $p$, $\dot{p}$ and $S$ are all of order $O(1)$ near the light horizon. Then, for a small $\rho$, the only large term in Eq.~(\ref{eq:EOM}) would be $p/\rho$, which cannot be balanced. Nevertheless, there are two scenarios to balance the equation in higher dimensions: Either 1D-like hopping occurs at some large value $\rho_*$, or $p$ goes to zero sufficiently fast at small $\rho$. The second scenario may be grossly approximated by the first in the limit $\rho_*\rightarrow 0$.

Now, let us find an approximation to the hopping solutions. Since the solution is continuous, there exist some $\rho_*$ near which $q$ crosses $-1$. The goal is to find equations that determine the unknown $\rho_*$. 
Suppose the width of the crossing layer $\Delta\rho\ll \rho_*$, then we can expand $\rho=\rho_*+x$ and $q=-1+y$. To lowest order, Eq.~(\ref{eq:EOM}) becomes $\ddot{y}+(D-1)\dot{y}/\rho_*-y\simeq S_*+\dot{S}_*x$, where $S_*$ and $\dot{S}_*$ are the source term and its derivative evaluated at $\rho_*$.  
For this linearized equation, the homogeneous problem is solved by $y\propto\exp(\omega x)$, and the eigenvalues are $\omega_\pm=\pm\sqrt{\eta^2+1}-\eta$, where $\eta=(D-1)/2\rho_*\ll1$. The general solution to the inhomogeneous problem is given by  $y=-S_*-(2\eta+x)\dot{S}_* + C_+\exp(\omega_+x)+ C_-\exp(\omega_-x)$, where $C_\pm$ will be determined by asymptotic matching.

When $\rho_*\gg1$, the hopping solution is constituted of four layers: the lower adiabatic layer, the crossing layer, the 1D hopping layer, and the upper adiabatic layer. 
To match the lower adiabatic and crossing layers, notice that $\omega_+>0$ and $\omega_-<0$, so we need $C_-=0$ to avoid the exponentially divergent behavior when $x\rightarrow-\infty$. 
Next, to match the crossing and 1D hopping layers, we need an exponential approximation to Eq.~(\ref{eq:1D_hopping}). Notice that when $q_\infty\sim1$, we have $q_c\sim-1$, $k\sim 1/2$ and $\theta\gg 1$. Then, in the intermediate regime where $kx\gg1$ and $\exp(kx)\ll\theta$, we can approximate 
$q_h(x)\simeq -1+ \exp(2kx)/\theta$.
Comparing this with the expression for $y$, in the crossing layer 
\begin{equation}
\label{eq:q_crossing}
q_{\text{cross}}\simeq-1-S_* -(2\eta+x)\dot{S}_* + \frac{e^{2kx}}{\theta}.
\end{equation}
Matching the exponents requires $\omega_+=2k$. Notice that $\omega_+$ depends on $\rho_*$ through $\eta$, while $k$ also depends on $\rho_*$ through $q_\infty$.
To lowest order, $q_\infty$ may be estimated from $S_*$. However, notice that the 1D hopping layer width may be large, so instead of expanding $S$ at $\rho_*$, estimating $q_\infty$ from $S$ at $\rho_0=\rho_*+\beta\rho_{1}$ gives a better approximation. Here, $\beta$ is a weighting parameter, and $\rho_{1}$ is given by Eq.~(\ref{eq:rc_1DRl}) after replacing $S_0$ with $S(\rho_0)$. Then, the equations that determine the crossing radius $\rho_*$ are
\begin{eqnarray}
\label{eq:q_rho}
\frac{1}{2} q_\infty(q_\infty^2-1)+S(\rho_0) &=& 0, \\
\label{eq:k_rho}
\frac{D\!-\!1}{\rho_*}\Big[\sqrt{1+\Big(\! \frac{2\rho_*}{D\!-\!1} \!\Big)^2}-1\Big] &=&\sqrt{2(3q_\infty^2-1)},
\end{eqnarray}
which require numerical root finding for a given $\beta$. For small $S_0$, the above equations have no real root, and hence there is no hopping FEP for weak sources. However, for intermediate $S_0$, the above equations have two real roots, which correspond to two hopping FEPs with different light horizon radii. Finally, for large $S_0$, there is a single real root, so a nonperturbative FEP always exists for strong sources.
The light horizon radius is $\rho_c=\rho_*+\rho_{1}$, 
which is usually dominated by the $\rho_*$, and the thickness of the hopping layer $\Delta\rho$ is 1D like.
Notice Eq.~(\ref{eq:q_rho}) is valid when $\rho_*$ is not much larger than $R$. Otherwise, $q_\infty$ should be estimated from the Yukawa potential, which is valid outside the source region.

Without relying on numerical root finding, the crossing radius may be estimated asymptotically. To lowest order, we can take the weighing parameter $\beta$ to be zero. Then, $q_\infty\simeq 1-S_*$ when $S_*\ll1$. Substituting this into Eq.~(\ref{eq:k_rho}), we see its right-hand side (RHS) is $\simeq2-3S_*$.  On the LHS, notice that the function is monotonously increasing with the asymptotic value $2$. Hence the solution to Eq.~(\ref{eq:k_rho}) satisfies $\rho_*\gg1$, which justifies this approximation to the hopping solution \textit{a posteriori}. 
In the limit $\rho_*\rightarrow\infty$, LHS$\simeq2-(D-1)/\rho_*$. Then, the equation becomes $(D-1)/\rho_*\simeq3S_*$, which can be written as $\xi\exp(-\xi^2)\simeq\epsilon$, where $\xi=\rho_*/R$ and $\epsilon=(D-1)/(3RS_0)$ is a small parameter. 
It is clear that when $\epsilon\ll1$, the transcendental equation has two real roots. The larger root may be approximated as $\xi\simeq(2\mu^3-1)/(2\mu^2-1)\sim\mu$, where $\mu=\sqrt{\ln(1/\epsilon)}$ gives the outer $\rho_*$ in Eq.~(\ref{eq:rho+}).
On the other hand, the smaller root may be approximated as $\xi\simeq\epsilon[\exp(\epsilon^2)-2\epsilon^2]/(1-2\epsilon^2)\sim\epsilon$, which gives the inner $\rho_*$ in Eq.~(\ref{eq:rho-}). Notice that although $\epsilon$ is small, $\rho_*\simeq\epsilon R$ may still satisfy $\rho_*\gg1$ because $R$ is larger, which is required for the approximations to be valid.

The conditions whereby two real roots exist give the lower and upper boundaries of the 1210 phase.
Notice that the maximum of $\xi\exp(-\xi^2)$ is $1/\sqrt{2e}$. Therefore, when $S_0$ is too small such that $\epsilon$ exceeds the maximum, the two solutions no longer exist. The condition $\epsilon=1/\sqrt{2e}$ marks the 1010-1210 phase boundary, which gives Eq.~(\ref{eq:A_1010^1210}). 
In the opposite limit, when $S_0$ is too large, $q_\infty$ decreases towards $1/\sqrt{3}$, in which case the RHS of Eq.~(\ref{eq:k_rho}) goes to zero. Then, the smaller root decreases towards $\rho_*=0$. In this case $S_*$ becomes $S_0$, and the critical value $S_0=1/3\sqrt{3}$ 
gives Eq.~(\ref{eq:A_2110^0110}) as the upper boundary of the 1210 phase. 
Notice that in the 1210 phase, $S_0\lesssim0.2$ is small, so the assumption that $S_*\ll1$ is always justified.
However, when $\rho_*\rightarrow 0$, 1D hopping is no longer a good approximation. As $A$ increases beyond the upper boundary of the 1210 phase, the inner hopping solution moves upwards and merges with the upper adiabatic solution, so a narrow 2110 phase appears before the system reaches the 0110 phase.

Finally, using the approximate crossing radius, the total energy of hopping FEPs can be estimated. 
Since the crossing layer is narrow, its energy contribution is subdominant, and the total energy $H_\pm\simeq H_a^-(\rho\!<\!\rho_0) +  H_a^+(\rho\!>\!\rho_0) + H_0$. 
The first term is the contribution from the lower adiabatic layer. Following Sec.~\ref{sec:1D:constant}, its energy density $\mathcal{E}\simeq -S$. Integrating $\mathcal{E}$ in $D$-dimensional space from $\rho=0$ to $\rho=\rho_0$, $H_a^-(\rho\!<\!\rho_0)\simeq-(A/2\pi^{D/2})\gamma(D/2,\rho_0^2/R^2)$, where $\gamma(a,z)=\int_0^zdt\; t^{a-1}\exp(-t)$ 
is the lower incomplete gamma function. 
Similarly, the second term is the contribution from the upper adiabatic layer. Integrating $\mathcal{E}\simeq S$ from $\rho=\rho_0$ to $\rho=\infty$, $H_a^+(\rho\!>\!\rho_0)\simeq(A/2\pi^{D/2})\Gamma(D/2,\rho_0^2/R^2)$, where $\Gamma(a,z)=\int_z^\infty dt\; t^{a-1}\exp(-t)$ 
is the upper incomplete gamma function. 
Finally, the third term $H_0$ is the contribution from the 1D hopping layer. Integrating the 1D energy density, 
$H_0=\int d\rho\;\rho^{D-1}\mathcal{E}_h$ now contains a surface contribution. Since the hopping width $\Delta\rho\ll\rho_*$, we can approximate $H_0\simeq \rho_0^{D-1} H_h$, where $H_h$ is given by Eq.~(\ref{eq:energy_hopping1D}). 
Then, summing the three contributions, the total energy of hopping FEPs is
\begin{equation}
\label{eq:energy_hopping}
H_\pm\simeq \frac{2}{3} \rho_0^{D-1} + \frac{A}{\pi^{D/2}}\Big[\Gamma\Big(\frac{D}{2},\frac{\rho_0^2}{R^2}\Big) -\frac{1}{2} \Gamma\Big(\frac{D}{2}\Big)\Big].
\end{equation}
To see how the energy depends on the source strength, 
$3(H_\pm-H_a^+)/R^{D-1}\simeq2\xi^{D-1}-[(D-1)/\epsilon]\gamma(D/2,\xi^2)$ may be regarded as a function of $\epsilon=(D-1)/(3RS_0)$ if $\xi=\rho_0/R$ is approximated using Eqs.~(\ref{eq:rho+}) and (\ref{eq:rho-}).
Then, above the 1010-1210 phase boundary, it is not difficult to see that $H_a^- < H_+< H_-$, which means that the ground state is always the adiabatic $q_a^-$ and the FEP with a larger light horizon always has lower energy. 
Moreover, using $\gamma(a,z)\simeq z^a /a$ when $z\rightarrow0$, $H_--H_a^+\simeq2(R\epsilon)^{D-1}/3D>0$. In other words, the highest-energy state is always the nonperturbative $q_-$. 
However, the ordering of $H_a^+$ and $H_+$ depends on the source strength: 
$H_a^+<H_+$ if and only if $S_0<S_c$, where $S_c$ may be estimated from $2\mu^{D-1}\simeq(D-1)\exp(\mu^2)\gamma(D/2,\mu^2)$.
Solving for $\mu^2=\ln(1/\epsilon)$, the asymptotic roots can be written in terms of 
$\epsilon_c(D=2)\simeq\exp[2/(1-e)]\approx0.31$, and 
$\epsilon_c(D=3)\simeq\exp[1/(1-\varsigma)]\approx0.38$, 
where $\varsigma=e\sqrt{\pi} \text{erf}(1)/2$. 
Notice that in the 1210 phase 
$\epsilon<1/\sqrt{2e}\approx0.43$, 
so level crossings occur within the 1210 phase and near its lower boundary. In other words, not far above the 1010-1210 boundary, the nonperturbative $q_+$ 
becomes the first excited state, in contrast to what happens in 1D where the adiabatic $q_a^+$ is always the first excited state.

\begin{figure}[t]
	\centering
	\includegraphics[width=0.48\textwidth]{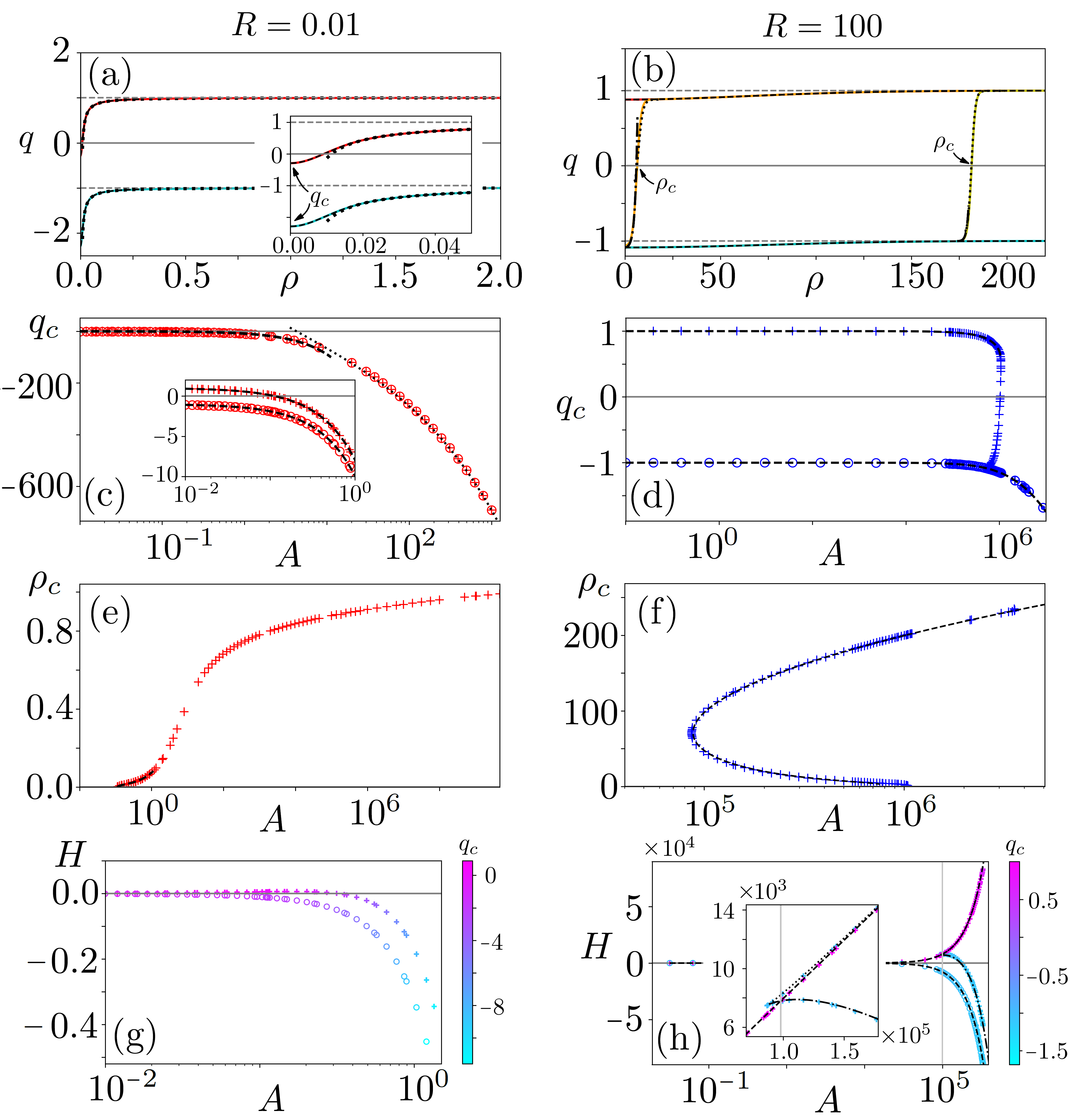}
	\caption{Example finite-energy potentials (FEPs), critical initial condition $q_c$, light horizon radius $\rho_c$, and normalized energy $H$ at $R=0.01$ and $R=100$ in 3D. Whenever applicable, asymptotic expressions (black) match numerical results (colored). 
	(a) Two FEPs exist when $A\approx0.14\gtrsim A_{1010}^{0110}$. The asymptotic solutions are constituted of the inner [Eq.~(\ref{eq:q_in_3D}), dashed] and outer [Eq.~(\ref{eq:q_out_3D}), dotted] layers.
	(b) Four FEPs exist when $A\approx5.5\times10^5$ is in the 1210 phase. The hopping solutions tunnel between the adiabatic solutions [Eq.~(\ref{eq:adiabatic}), dashed] via the crossing [Eq.~(\ref{eq:q_crossing}), dash-dot] and hopping [Eq.~(\ref{eq:1D_hopping}), dotted] layers.
	(c) When $R\ll1$, $q_c$ decreases linearly with $A$ [Eq.~(\ref{eq:qc_3D}), dashed] for $A\lesssim O(4\pi^{3/2} R)$, while $q_c$ scales with $A^{1/3}$ (dotted) in the nonperturbative regime $\frac{1}{2}q^3+S\simeq0$. 
	(d) When $R\gg1$, $q_c$ for the adiabatic solutions is given by the largest and smallest real roots of $\frac{1}{2}q_c(q_c^2-1)+S_0\simeq0$ (dashed).
	(e) Near the phase boundary, $\rho_c\rightarrow R\sqrt{3(1-A_{1010}^{0110}/A)}$ (dashed) may be determined from the inner layer $q_\text{in}(\rho_c)=0$. 
	(f) In the 1210 phase, $\rho_c$ has two possible values correspond to $\rho_\pm$ [Eqs.~(\ref{eq:rho+}) and (\ref{eq:rho-})]. The crossing radii are estimated as roots of Eqs.~(\ref{eq:q_rho}) and (\ref{eq:k_rho}) with $\beta=0.95$ (dashed).
	(g) In the perturbative regime, $H\sim O(A)$ and $q_-$ and $q_+$ are the ground and excited states, respectively. 
	(h) In the 1210 phase, the energy levels of hopping [Eq.~(\ref{eq:energy_hopping}), dash-dot] and adiabatic [Eq.~(\ref{eq:energy_adiabatic}), dashed] FEPs cross near $A=(2\pi^{3/2}/3\epsilon)R^2\approx9.8 R^2$ (gray). 
	}
	\label{fig:asymptotic}
\end{figure}

\subsection{Numerical results \label{sec:DD:numeric}}
Since the behaviors in 2D and 3D are qualitatively similar, to keep the discussion compact, parts of the numerical results will only be shown for 2D, while complementary results will be shown for 3D, which are obtained using the following method. 
First, for a given $R$ and $A$ with $p_0=0$, the initial condition $q_0$ is scanned to search for boundaries across which numerical solutions to Eq.~(\ref{eq:EOM}) transition from divergent to oscillatory. On each boundary, the critical initial value $q_c$, which is determined using bisection, gives rise to a FEP. 
Notice that at any finite precision, since the VEVs are saddle points of the dynamical system, numerical solutions always oscillate or diverge at finite $\rho_f$, which goes to infinity only when $q_0\rightarrow q_c$. For results reported here, the precision $|q_0-q_c|$ is set to target for $\rho_f>5\max(1,R)$, which ensures that both the Compton and source scales are captured at a manageable numerical cost.
For each FEP, its energy is estimated by numerically integrating Eq.~(\ref{eq:energy_density}) from $\rho=0$ to $\rho=\tilde{\rho}_f<\rho_f$, where $\mathcal{E}(\tilde{\rho}_f)$ is the first local minimum after $H$ has reached a plateau.  
Moreover, when the FEP exhibits a light horizon, its radius $\rho_c$ is estimated from the numerical solution. 
Second, using $q_c$ and $\rho_c$ as two order parameters, phases in the $R$-$A$ parameter space are identified. At a given $R$, the order parameters are computed as multi-valued functions of $A$. Then, bisections are used to determine phase boundaries $A_c$, across which the number of $q_c$ and $\rho_c$ values jumps. Finally, $R$ is scanned to map out the $R$-$A$ parameter space. Since the behaviors for $R\ll1$ and $R\gg1$ are known asymptotically, which match numerical results (Fig.~\ref{fig:asymptotic}), scanning $R$ between 0.01 and 100 is sufficient to characterize the entire $R$-$A$ parameter space.
For $R\le10$, numerical solutions with a decimal precision of 28 is usually sufficient, but for larger $R$, the decimal precision is set to 125 or higher in order to resolve the 1210 phase, which shows up as a narrow secondary island.

\begin{figure}[b]
	\centering
	\includegraphics[width=0.48\textwidth]{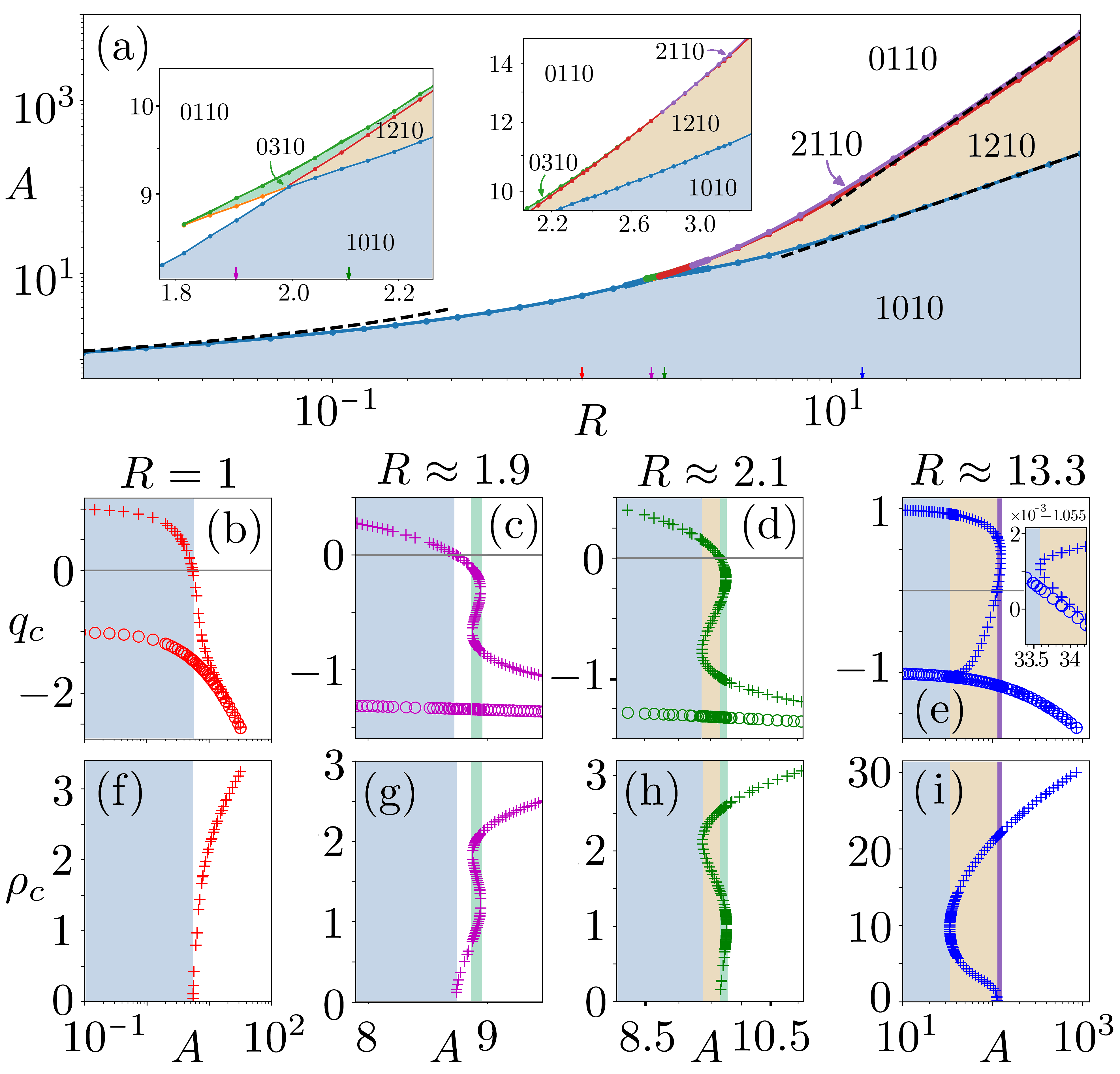}
	\caption{(a) The 2D $R$-$A$ phase diagram is constituted of five phases, whose asymptotic boundaries (dashed black) match numerical results (colored dots). 
	(b) $R=1$ shows an example of $q_c$ at small $R$ near the 1010 (gray) and 0110 (white) phase boundary. In this regime, there are always two critical initial conditions, one of which leads to a finite-energy-potential (FEP) with $q(\infty)=+1$ (``+" marker) while the other leads to a FEP with $q(\infty)=-1$ (``$\circ$" marker). 
	(c) $R\approx1.9$ and (d) $R\approx2.1$ show two examples at intermediate $R$, where the 0310 (green) and 1210 (bisque) phases emerge. In these phases, there are four $q_c$ values for each $A$. 
	(e) $R\approx13.3$ shows an example at large $R$, where the ``+" branch folds towards the ``$\circ$" branch on the left while bulges up on the right, giving rise to a narrow 2110 phase (purple). 
	(f) Whenever $q_c<0$, the FEP with $q(\infty)=1$ exhibits a light horizon with radius $\rho_c$. For small $R$, $\rho_c$ is a single valued function of $A$ and the derivative $\rho'_c>0$ at $A_c$. 
	(g) At intermediate $R$, $\rho_c$ folds back and becomes a triple-valued function in the 0310 phase. 
	(h) At larger $R$, $\rho_c$ folds further back and is double valued in the 1210 phase. 
	(i) When $R\gg1$, the derivative $\rho'_c(A_c)<0$ changes sign, and $\rho_c(A)$ transitions from a double-valued to a single-valued function above the 1210-2110 phase boundary.		
	}
	\label{fig:2D}
\end{figure}

The 2D phase diagram and examples of the underlying order parameters are shown in Fig.~\ref{fig:2D}. The $R$-$A$ parameter space (Fig.~\ref{fig:2D}a) is constituted of three major regions: the 0110 (white), 1010 (gray), and 1210 (bisque) phases. A fourth 2110 phase appears between the 1210 and 0110 phases as a narrow stripe (purple). The numerical phase boundaries (colored dots) approach asymptotic results (dashed black) when $R\rightarrow 0$ and $R\rightarrow\infty$.
In the intermediate region $R\sim1$, the phase diagram is constituted of an additional 0310 phase (green), which is bordered with other phases via a double point and two quadruple points (Fig.~\ref{fig:2D}a, insets). On the left is the double point, where a wedge of the 0310 phase cuts into the 0110 phase. In the middle is a quadruple point where the 0110, 0310, 1210, and 1010 phases meet. On the right is another quadruple point, where the 0310 phase attenuates while the 2110 phase emerges between the 0110 and 1210 phases.
The five phases have distinct features, as characterized by the critical initial condition $q_c$ (Fig.~\ref{fig:2D}b-\ref{fig:2D}e) and the light horizon radius $\rho_c$ (Fig.~\ref{fig:2D}f-\ref{fig:2D}i).  
In the 1010 phase, there are two values of $q_c$ at each $A$. The upper (lower) $q_c$ leads to a FEP with $q(\infty)=+1 (-1)$, and neither FEP has a light horizon.
In comparison, in the 0110 phase, there are also two FEPs, but now the upper FEP exhibits a light horizon. 
While the 1010-0110 phase transition only changes the number of light horizons, the 0110-0310 phase transitions also change the number of FEPs: In the 0310 phase, there are four FEPs, three of them exhibit light horizons. Notice that at each $A$, there is no longer a unique solution for a given boundary condition. 
The situation is similar in the 1210 phase, except that only two of the four FEPs have light horizons.  
Finally, in the 2110 phase, there are again four FEPs and one of them has a light horizon. 
Notice that at large $R$, the two FEPs in the 0110 phase have infinitesimally close $q_c$ when $A\rightarrow\infty$. In other words, two potentials with hardly distinguishable initial conditions bifurcate towards disparate boundary values.

\begin{figure}[b]
	\centering
	\includegraphics[width=0.48\textwidth]{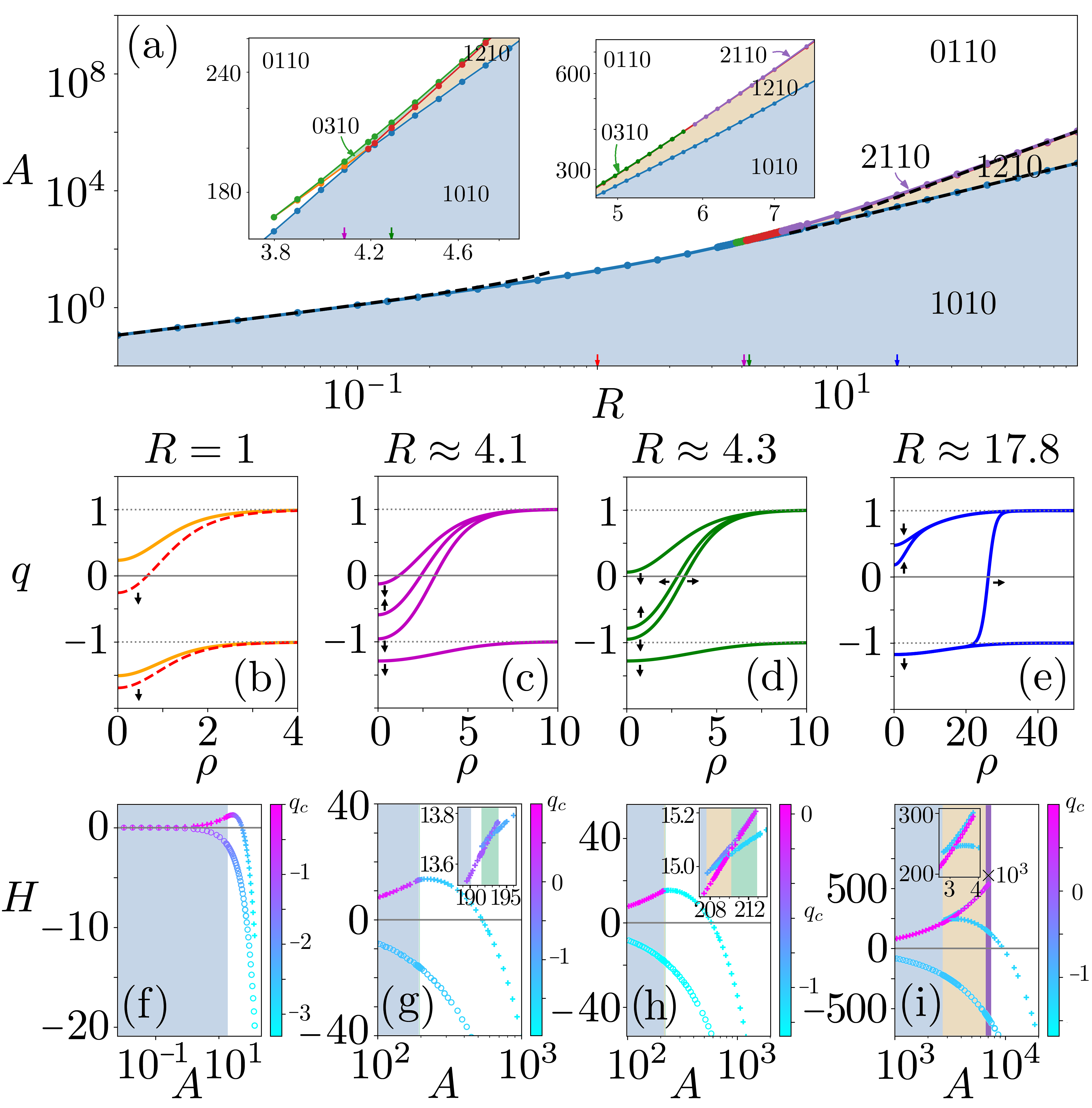}
	\caption{(a) The 3D $R$-$A$ phase diagram is also constituted of five phases, whose asymptotic boundaries (dashed black) match numerical results (colored dots). 
		(b) As shown by example solutions at $R=1$, in the 1010 ($A\approx 14.9$, orange) and 0110 ($A=21.6$, red) phases, there is a unique finite-energy-potential (FEP) with $q(\infty)=\pm1$. 
		(c) At intermediate $R\approx4.1$, the 0310 phase emerges, where three FEPs have $q(\infty)=+1$ when $A\approx192.7$. 
		(d) The situation is similar at larger $R\approx4.3$ in the 1210 phase, except that the upper FEP no longer exhibits a light horizon when $A\approx208.0$. 
		(e) At large $R\approx17.8$ and $A\approx 7005.1$, the 2110 phase appears. The two upper FEPs merge whereas the two lower FEPs split as $\rho$ increases. 
		In these figures, black arrows indicate the directions towards which the FEPs move when $A$ increases. 
		(f) Energy of the lower FEP $H^-$ decreases with $A$, whereas $H^+$ is not monotonous.
		(g) At intermediate $R$, $H^+$ folds up near the maximum, and two additional energy levels arise in the 0310 phase (green).
		(h) The four energy levels extend further in the 1210 phase (bisque) at large $R$.
		(i) For even larger $R$, the 0310 phase is replaced by the 2110 phase (purple). 
	}
	\label{fig:3D}
\end{figure}

The 3D phase diagram, 
together with example FEPs and the normalized energy are shown in Fig.~\ref{fig:3D}. 
For small $R$ (Fig.~\ref{fig:3D}b), there is a unique FEP for a given boundary condition at each $A$.  For example, at $A\approx14.9$ (solid orange), the potentials are noticeably depleted from the VEVs on the Compton scale; At larger $A=21.6$ (dashed red), the potentials are depleted further, and the upper FEP crosses zero, giving rise to a light horizon. 
While the energy for the lower branch $H^-$ monotonously decreases with $A$, the energy for the upper branch $H^+$ has a maximum near $A\approx28$ in the 0110 phase (Fig.~\ref{fig:3D}f).
The decreasing behavior is expected, because at large $A$ where $S\gg1$, the dominant balance of Eq.~(\ref{eq:EOM}) is $\frac{1}{2}q^3+S\simeq0$, so the solution $q\simeq-(2S)^{1/3}$ becomes universal. Then, the energy density [Eq.~(\ref{eq:energy_density})] is dominated by the $Sq$ term, which is negative and decreases with $A$. It is worth noting that the energy splitting $H^+-H^-$ always increases with $A$.
While these large-scale trends of $H^\pm$ remain the same for all cases, extra details develop near the maximum of $H^+$ for larger $R$. For example, at $R\approx4.1$ (Fig.~\ref{fig:3D}g), the $H^+$ branch folds back sharply and two additional energy levels emerge in the 0310 phase. In this phase, three FEPs have the same boundary value $q(\infty)=1$, and example solutions are shown in Fig.~\ref{fig:3D}c when $A\approx192.7$. 
The FEPs are similar in the 1210 phase when $R\approx4.3$ and $A\approx208.0$ (Fig.~\ref{fig:3D}d), except that the upper FEP no longer exhibits a light horizon. Near the 1010-1210 boundary, the upper adiabatic FEP is the first excited state and the outer hopping FEP is the second excited state (Fig.~\ref{fig:3D}h, inset). However, the energy ordering is switched at larger $A$. 
This switching behavior remains for larger sources, for example at $R\approx17.8$ (Fig.~\ref{fig:3D}i), except that a shrinking 0310 phase is replaced by an expanding 2110 phase, in which the inner light horizon disappears.
In the 2110 phase, as shown in Fig.~\ref{fig:3D}e where $A\approx7005.1$, the two upper FEPs merge whereas the two lower FEPs split when $\rho$ increases. After $A$ surpasses the 2110-0110 phase boundary, the two upper FEPs annihilate while the two lower FEPs split at ever larger radius.

\section{Discussion \label{sec:discussion}}
While quantitative results in this paper are specific to a real scalar field with a Maxican-hat self nonlinearity under the influence of a Gaussian source, qualitative features of nonperturbative potentials are likely more general, provided that the two key ingredients, namely, the nonlinearity and the source, are present.
First, for multiple FEPs to exist, the nonlinearity should allow more than one local minimum at which a vacuum field configuration is stable. A stable minimum against temporal fluctuations is a saddle point for static spatial configurations, which means that any slight deviation from the VEV will lead to runaway distortions that allow the field to hop spatially from one VEV to another.
Second, for the hopping to occur at a finite energy cost, an external source is often necessary. This is because different VEV may be associated with different topological charges, so tunneling between VEVs may require breaking topological invariance, which is possible with an external source.
The source term, which may seem artificial, arises when one focuses on a subsystem of a bigger problem. For example, when studying electronic structures of materials, atomic nuclei are usually prescribed, which provide external Coulomb potentials for electrons. 
As another example, in fluid dynamics, a boundary foreign to the flow set a scale for the Reynolds number, which determines whether the flow is in the laminar, vortex, or turbulent phases.
Here, the external source, which may originate from other particles and interactions, allow a FEP to tunnel between VEVs by providing the necessary activation. 
Regardless of its exact functional form, the source has certain characteristic size and strength. Since these parameters introduce additional scales, interesting phenomena can occur in the presence of external sources. For small sources, they give instantaneous kicks, while for large sources, they adiabatically deform the VEVs, which can spill over into another local minimum when the deformation is strong enough. A remarkable consequence is that FEPs can vary on scales that are much larger than the Compton wavelength.

To show that qualitative features of nonperturbative potentials are preserved for other source profiles, let us use a Lorentzian source $S=S_0/(1+\rho^2/R^2)$ as an example. Consider the 1210 phase in $D\ge 1$ dimensions when $R\gg1$. Then, the hopping potentials are constituted of the four layers described in Sec.~\ref{sec:DD:constant}. To connect the layers, one again solves for Eqs.~(\ref{eq:q_rho}) and (\ref{eq:k_rho}), except that now the source profile is Lorentzian. Denoting $S_*$ the source strength at the crossing radius $\rho_*$, then in the limit $S_*\ll1$, the equations are well approximated by $(D-1)/\rho_*\simeq3S_*$, which can be written as $\xi^2+1\simeq\xi/\epsilon$, where $\xi=\rho_*/R$ and $\epsilon=(D-1)/(3RS_0)$.
The condition that real roots exist is $\epsilon\le1/2$, so the asymptotic 1010-1210 phase boundary is $S_c\simeq2(D-1)/(3R)$, which decreases linearly with $R$. 
Hence, up to some $O(1)$ constants with $S_0\sim A/R^D$, the phase boundary for a Lorentzian source is the same as that for a Gaussian source [Eq.~(\ref{eq:A_1010^1210})], which scales as $A_c\propto R^{D-1}$. 
Within the 1210 phase when $\epsilon\ll1$, the lowest-order asymptotic expressions for the crossing radii are $\xi_+\simeq1/\epsilon$ and $\xi_-\simeq\epsilon$. In other words, $\rho_+\simeq(3S_0R^2)/(D-1)\gg R$ and $\rho_-\simeq (D-1)/(3S_0)\ll R$. Although the exact dependencies on $R$ and $S_0$ are different from Eqs.~(\ref{eq:rho+}) and (\ref{eq:rho-}), we see that both Lorentzian and Gaussian sources allow nonperturbative hopping potentials, whose scales are potentially well beyond the Compton wavelength.

\begin{table*}
	\caption{\label{tab:para}Example sources for the Higgs field in 3D. 
		The source distribution is assumed to be a spherical Gaussian, which may be inaccurate but sufficient for order-of-magnitude estimations. The estimates are made using asymptotic expressions whenever applicable and numerical solutions otherwise.
		In most cases, the Higgs is in the perturbative 1010 phase, and the ratio $A/A_c$ indicates the distance from the phase boundary, where $A_c=A_{1010}^{1210}$ ($A_c=A_{1010}^{0110}$) for $R>1$ ($R<1$). $r_c$ is the light horizon radius, which is present only in nonperturbative phases. $E_0$ is the energy of the ground-state configuration, and $\Delta E$ is the gap between the ground and the first excited states. 
		As reference scales, contemporary inertial confinement fusion experiments input $\sim10^6$ J; Hydrogen bombs release $\sim10^{16}$ J; supernovae release $\sim10^{10}L_\odot\cdot$yr, where $L_\odot$ is the solar luminosity. }
	\begin{ruledtabular}
		\begin{tabular}{c cc cc c c c cc}
			Source  & size & $R$ & mass & $A$ & $A/A_c$ & phase & $r_c$ & $-E_0$ & $\Delta E$ \\
			\hline
			\rule{0pt}{2.5ex}   
			hypothetical fermion
			& $0.2$ am \footnote{The scales of elementary fermions are estimated using their Compton wavelengths.}
			& 0.1 & $10^3$ GeV & $10^0$ & $10^0$ & 0110 & $0.2$ am \footnote{For heavier fermions, the light horizon radius is on the order of the Higgs Compton wavelength \mbox{$\lambdabar_c\sim 1$ am}.}
			& $10^3$ GeV & $10^3$ GeV \\
			top quark & $1$ am \footnotemark[1] & 0.7 & 173 GeV & $10^{-1}$ & $10^{-2}$ & 1010 & & $10^2$ GeV  & $10^2$ GeV \\
			$\phantom{}^{12}\text{C}$ & 3 fm & $2\times10^3$ & 11.3 GeV & $10^{-4}$ & $10^{-11}$ & 1010 & & $10^{-1}$ GeV & $10^{-1}$ GeV \\
			$\phantom{}^{197}\text{Au}$ & 8 fm & $5\times10^3$ & 185 GeV & $10^{-3}$ & $10^{-11}$ & 1010 & & $10^{0}$ GeV & $10^{0}$ GeV \\
			electron & 0.4 pm \footnotemark[1] & $2\times10^5$ & 0.511 MeV & $10^{-6}$ & $10^{-18}$ & 1010 & & $10^{0}$ MeV & $10^{0}$ MeV \\
			\hline
			\rule{0pt}{2.5ex}  
			fusion stagnation & 50 $\mu$m & $3\times 10^{13}$ & 1 mg & $10^{16}$ & $10^{-12}$ & 1010 & & $10^{9}$ J & $10^{9}$ J\\  
			uranium ball & 1 m & $6\times10^{17}$ & $80$ t & $10^{27}$ & $10^{-10}$ & 1010 & & $10^{20}$ J & $10^{20}$ J \\
			\hline
			\rule{0pt}{2.5ex}  
			neutron star & 10 km & $10^{22}$ & $1.5M_{\odot}$ & $10^{52}$ & $10^{8}$ & 1210 & 40 $\mu$m, 50 km & $10^{11}\;L_\odot\cdot$yr & $10^{4}\;L_\odot\cdot$yr\\
			earth & $6\times10^3$ km & $4\times10^{24}$ & $6\times10^{24}$ kg & $10^{47}$ & $10^{-3}$ & 1010 & & $10^{5}\;L_\odot\cdot$yr & $10^{6}\;L_\odot\cdot$yr \\
			white dwarf & $10^4$ km & $10^{25}$ & $0.6M_{\odot}$ & $10^{52}$ & $10^{2}$ & 1210 & 70 km, $2\!\times\!10^4$ km & $10^{11}\;L_\odot\cdot$yr & $10^{10}\;L_\odot\cdot$yr \\
			Jupiter & $7\times10^4$ km & $4\times10^{25}$ & $2\times10^{27}$ kg & $10^{49}$ & $10^{-3}$ & 1010 & & $10^{8}\;L_\odot\cdot$yr & $10^{8}\;L_\odot\cdot$yr \\				
			Sun & $R_{\odot}$ & $4\times10^{26}$ & $M_{\odot}$ & $10^{52}$ & $10^{-2}$ & 1010 & & $10^{11}\;L_\odot\cdot$yr & $10^{11}\;L_\odot\cdot$yr \\
			supergiant & $500R_{\odot}$ & $10^{29}$ & $15M_{\odot}$ & $10^{53}$ & $10^{-6}$ & 1010 & & $10^{12}\;L_\odot\cdot$yr & $10^{12}\;L_\odot\cdot$yr \\
			\hline
			\rule{0pt}{2.5ex}  
			dwarf galaxy halo \footnote{Dark matter and baryonic matter are assumed to have comparable coupling with the Higgs. The size and mass are virial.}
			& 25 kpc & $10^{38}$ & $10^9 M_{\odot}$ & $10^{61}$ & $10^{-17}$ & 1010 & & $10^{20}\;L_\odot\cdot$yr & $10^{20}\;L_\odot\cdot$yr \\
			Milky Way halo\footnotemark[3] & 280 kpc & $10^{40}$ & $10^{12} M_{\odot}$ & $10^{64}$ & $10^{-16}$ & 1010 & & $10^{23}\;L_\odot\cdot$yr & $10^{23}\;L_\odot\cdot$yr \\
		\end{tabular}
	\end{ruledtabular}
\end{table*}

The nonperturbative potentials may be realized in many physical systems. For example, in condensed matter systems, impurities may serve as source terms, and the effective field theory may contain nonlinearities that allow for spontaneous symmetry breaking. Notably, in the Ginzburg-Landau model of superconductivity \cite{Ginzburg50}, the nonlinearity is exactly of the Maxican-hat type below the critical temperature, albeit that the effective field of Cooper pairs is a complex scalar field, which may allow additional vortex-like nonperturbative states. 
Changes to the Cooper-pair field $\phi$ may be induced by an isolated impurity or simply an electrode, whose size and strength may be controlled experimentally. Here, the size should be compared to the Compton wavelength $\lambda_*$ associated with the effective mass $m_*$.
For small sources in $D\ge2$ dimensions, the field configuration is perhaps not very interesting: The local disturbance of $\phi$ decays exponentially on the scale of $\lambda_*$. However, interesting phenomenon may occur, for example, in a 1D nanowire \cite{Altomare2013one}: When the impurity strength is below a critical value, there exists an excited state of $\phi$ such that the Cooper pairs are depleted at a distance $r_c\gg\lambda_*$ away from the source, and $r_c$ increases when the impurity strength diminishes. 
Additionally, interesting phenomena may occur in all dimensions when the source is of larger sizes. In this case, as one adjusts the source strength, horizons and vortices may be turned on or off. Moreover, since each additional FEP is associated with an extra energy level, properties of the material, such as the specific heat, susceptibility, and permittivity, may experience accompanying phase transitions. In other words, one may be able to add or remove energy levels discontinuously by adjusting the source parameters.
Notice that such phase transitions are controlled by a localized source within the material, without the need of changing extrinsic global parameters such as temperature and magnetic fields.

Another important example is the Higgs field. In the standard model of particle physics, the Higgs field originates from a SU(2) doublet, whose nonlinear self-interaction is exactly of the Maxican-hat type \cite{Higgs64}. 
Although there are additional effects due to the non-abelian group \cite{Dashen74nonabelian}, results for the real scalar field may be used to give order-of-magnitude estimations regarding when nonperturbative effects become important for the Higgs field and its beyond-standard-model extensions \cite{Witzel2019review}. 
Taking parameters for the Higgs field \cite{Particle2020review}, \mbox{$v\approx 246$ GeV} and $m\approx$ \mbox{125 GeV}, which corresponds to a Compton wavelength of $\lambdabar_c=\hbar/mc\approx$\mbox{$1.6\times10^{-18}$ m}.
In 3D, the normalized source strength is $A=\alpha m/v$. To see how $\alpha$ might be interpreted, notice that elementary fermions couple to the Higgs field via $\mathcal{L}_f=-m_f \bar{f}f\phi/v$ where $m_f$ is the fermion mass. Comparing $\mathcal{L}_f$ to Eq.~(\ref{eq:energy_density}), the source term $\mathscr{S}=(m_f/v)\langle \bar{f}f\rangle$, where $\langle \bar{f}f\rangle$ is related to the particle number density. Then, $\alpha\sim(m_f/v)N$, 
where $N$ is the number of particles that constitute the source. 
First, consider elementary particles, for which $N=1$. For low-mass fermions such as electrons, $A\sim mm_f/v^2\ll1$ while $R\sim m/m_f\gg1$. Then, from Fig.~\ref{fig:3D}, we see $\phi$ is in the 1010 phase, where the unique potential is well approximated by the Yukawa potential. 
On the other hand, for heavy fermions such that $R\lesssim1$, the 1010-0110 phase boundary is given by Eq.~(\ref{eq:Ac_Rs_3D}). To lowest order, the critical mass is $m_c^2=2\pi^{3/2}v^2$, and to the next order $m_c\approx3.57v\approx$ \mbox{878 GeV}. Notice that the top quark mass is $m_f\approx$ \mbox{173 GeV}, so all known elementary fermions are in the 1010 phase and the potentials are always perturbative. However, suppose hypothetical fermions with mass \mbox{$m_f\gtrsim10^3$ GeV} exist, then they are in the nonperturbative 0110 phase, for which the Higgs field vanishes on a light horizon whose radius $r_c\sim\lambdabar_c$. 
Second, consider composite-particle sources, for which atomic nuclei are perhaps the most common examples \cite{Choppin2002radiochemistry}. Although nuclei do not directly couple with the Higgs, its constituent quarks do. Counting valance quarks only, $N\simeq3M/m_p$ where $M$ is the total mass of the source and $m_p$ is the proton mass. Then, $\alpha\sim3Mm_f/m_pv$, where $m_f$ may be estimated from the up and down quark masses $(m_u+m_d)/2$\mbox{$\approx3.45$ MeV}. The normalized source strength is then $A\sim2.3\times10^{-5} M_{\text{GeV}}\approx1.3\times10^{22} M_{\text{kg}}$, where $M_{\text{GeV}}$ and $M_{\text{kg}}$ are $M$ in units of GeV and kg, respectively. 
For a single nucleus, its size is $a\simeq a_0(M/m_p)^{1/3}$ where \mbox{$a_0\approx1.4$ fm}, so $R\sim9\times10^{2} M_{\text{GeV}}^{1/3}$ is large. We see all stable nuclei are well within the 1010 phase.
However, for an aggregated amount of matter, the situation becomes more interesting. Using Eq.~(\ref{eq:A_1010^1210}), the 1010-1210 phase boundary in 3D is $A_{1010}^{1210}\approx8.7 R^2\approx3.5\times10^{36} a_{\text{m}}^2$, where $a_{\text{m}}$ is the source size in units of meter. Then, the critical mass for reaching the 1210 phase is \mbox{$M_{1010}^{1210}\sim2.7\times10^{14} a_{\text{m}}^2$ kg}, which is reduced by an $O(1)$ factor if we also account for electrons. 
Since $M_{1010}^{1210}\propto a^2$, the Higgs field will always transitions into the 1210 phase for a sufficiently large source at a given density.
Further transitions into the 2110 and 0110 phases are formidable, because the requisite $A_c\approx1.1 R^3\approx2.7\times10^{53} a_{\text{m}}^3$ [Eq.~(\ref{eq:A_2110^0110})]. The corresponding \mbox{$M_c\sim2.1\times10^{31}a_{\text{m}}^3$ kg}, which means that the critical density far exceeds even the nuclear matter density $\sim10^{17}\; \text{kg/m}^3$ and is independent of the source size.
A number of source examples are listed in Table.~\ref{tab:para}, ranging from microscopic particles \cite{Choppin2002radiochemistry}, macroscopic objects \cite{Lindl1995development,Bethe1950hydrogen}, astrophysical bodies \cite{Norton2002introduction}, to galaxies \cite{Revaz2018pushing,Posti2019mass}. In most cases, the Higgs is well within the perturbative 1010 phase. However, for astrophysical bodies such as white dwarfs and neutron stars, the Higgs is potentially in the nonperturbative 1210 phase where light horizons exist.
Near the light horizons, the Higgs expectation value is depleted, so the mass of all elementary particles are reduced proportionally.
The resultant universal mass gradient mocks many effects of gravity \cite{Shi2019force}. For example, all objects accelerate equally towards the light horizon, and photons emitted from bound states are redshifted. 
Moreover, the kinematics of particle interactions is altered. In particular, since the $W$-boson mass is reduced, the Fermi constant increases, which may change nuclear states and their reaction rates.

In summary, in the presence of symmetry-breaking nonlinearities and external sources, finite-energy configurations of a massive scalar field are not always unique for given boundary conditions. Apart from the well known Yukawa potential, the scalar field may exhibit additional nonperturbative configurations, such as the hopping potentials, which vary on scales that are much larger than the Compton wavelength. Exactly what static potentials are allowed depends on the source size and strength, and the phase diagrams in one, two, and three spatial dimensions are mapped out completely for an isolated Gaussian source by combining asymptotic and numerical solutions.
Nonperturbative potentials may have observable consequences in many physical systems. Dynamical effects when sources move and fields fluctuate remain to be explored in the future.

\section*{Data and code availability}	
The data underlying numerical results are available at \href{https://doi.org/10.5281/zenodo.5021378}{\url{https://doi.org/10.5281/zenodo.5021378}}.
The computer codes used to generate and plot the data are available at \href{https://github.com/seanYuanSHI/Phi4}{\url{https://github.com/seanYuanSHI/Phi4}}.

\begin{acknowledgments}
This work was performed under the auspices of the U.S. Department of Energy by Lawrence Livermore National Laboratory under Contract DE-AC52-07NA27344 and was supported by the Lawrence Fellowship through LLNL-LDRD Program under Project No. 19-ERD-038. 
\end{acknowledgments}


\begin{thebibliography}{24}%
	\makeatletter
	\providecommand \@ifxundefined [1]{%
		\@ifx{#1\undefined}
	}%
	\providecommand \@ifnum [1]{%
		\ifnum #1\expandafter \@firstoftwo
		\else \expandafter \@secondoftwo
		\fi
	}%
	\providecommand \@ifx [1]{%
		\ifx #1\expandafter \@firstoftwo
		\else \expandafter \@secondoftwo
		\fi
	}%
	\providecommand \natexlab [1]{#1}%
	\providecommand \enquote  [1]{``#1''}%
	\providecommand \bibnamefont  [1]{#1}%
	\providecommand \bibfnamefont [1]{#1}%
	\providecommand \citenamefont [1]{#1}%
	\providecommand \href@noop [0]{\@secondoftwo}%
	\providecommand \href [0]{\begingroup \@sanitize@url \@href}%
	\providecommand \@href[1]{\@@startlink{#1}\@@href}%
	\providecommand \@@href[1]{\endgroup#1\@@endlink}%
	\providecommand \@sanitize@url [0]{\catcode `\\12\catcode `\$12\catcode
		`\&12\catcode `\#12\catcode `\^12\catcode `\_12\catcode `\%12\relax}%
	\providecommand \@@startlink[1]{}%
	\providecommand \@@endlink[0]{}%
	\providecommand \url  [0]{\begingroup\@sanitize@url \@url }%
	\providecommand \@url [1]{\endgroup\@href {#1}{\urlprefix }}%
	\providecommand \urlprefix  [0]{URL }%
	\providecommand \Eprint [0]{\href }%
	\providecommand \doibase [0]{https://doi.org/}%
	\providecommand \selectlanguage [0]{\@gobble}%
	\providecommand \bibinfo  [0]{\@secondoftwo}%
	\providecommand \bibfield  [0]{\@secondoftwo}%
	\providecommand \translation [1]{[#1]}%
	\providecommand \BibitemOpen [0]{}%
	\providecommand \bibitemStop [0]{}%
	\providecommand \bibitemNoStop [0]{.\EOS\space}%
	\providecommand \EOS [0]{\spacefactor3000\relax}%
	\providecommand \BibitemShut  [1]{\csname bibitem#1\endcsname}%
	\let\auto@bib@innerbib\@empty
	\bibitem [{\citenamefont {Weinberg}(2012)}]{Weinberg2012classical}%
	\BibitemOpen
	\bibfield  {author} {\bibinfo {author} {\bibfnamefont {E.~J.}\ \bibnamefont
			{Weinberg}},\ }\href@noop {} {\emph {\bibinfo {title} {{Classical solutions
					in quantum field theory: Solitons and Instantons in High Energy Physics}}}}\
	(\bibinfo  {publisher} {Cambridge University Press},\ \bibinfo {year}
	{2012})\BibitemShut {NoStop}%
	\bibitem [{\citenamefont {Dashen}\ \emph
		{et~al.}(1974{\natexlab{a}})\citenamefont {Dashen}, \citenamefont
		{Hasslacher},\ and\ \citenamefont {Neveu}}]{Dashen74extended}%
	\BibitemOpen
	\bibfield  {author} {\bibinfo {author} {\bibfnamefont {R.~F.}\ \bibnamefont
			{Dashen}}, \bibinfo {author} {\bibfnamefont {B.}~\bibnamefont {Hasslacher}},\
		and\ \bibinfo {author} {\bibfnamefont {A.}~\bibnamefont {Neveu}},\ }\bibfield
	{title} {\bibinfo {title} {{Nonperturbative methods and extended-hadron
				models in field theory. II. Two-dimensional models and extended hadrons}},\
	}\href {https://doi.org/10.1103/PhysRevD.10.4130} {\bibfield  {journal}
		{\bibinfo  {journal} {Phys. Rev. D}\ }\textbf {\bibinfo {volume} {10}},\
		\bibinfo {pages} {4130} (\bibinfo {year} {1974}{\natexlab{a}})}\BibitemShut
	{NoStop}%
	\bibitem [{\citenamefont {Abrikosov}(1957)}]{Abrikosov1957}%
	\BibitemOpen
	\bibfield  {author} {\bibinfo {author} {\bibfnamefont {A.}~\bibnamefont
			{Abrikosov}},\ }\bibfield  {title} {\bibinfo {title} {The magnetic properties
			of superconducting alloys},\ }\href
	{https://doi.org/https://doi.org/10.1016/0022-3697(57)90083-5} {\bibfield
		{journal} {\bibinfo  {journal} {J. Phys. Chem. Solids}\ }\textbf {\bibinfo
			{volume} {2}},\ \bibinfo {pages} {199} (\bibinfo {year} {1957})}\BibitemShut
	{NoStop}%
	\bibitem [{\citenamefont {'t~Hooft}(1974)}]{Hooft1974magnetic}%
	\BibitemOpen
	\bibfield  {author} {\bibinfo {author} {\bibfnamefont {G.}~\bibnamefont
			{'t~Hooft}},\ }\bibfield  {title} {\bibinfo {title} {Magnetic monopoles in
			unified gauge theories},\ }\href@noop {} {\bibfield  {journal} {\bibinfo
			{journal} {Nucl. Phys. B}\ }\textbf {\bibinfo {volume} {79}},\ \bibinfo
		{pages} {276} (\bibinfo {year} {1974})}\BibitemShut {NoStop}%
	\bibitem [{\citenamefont {Polyakov}(1974)}]{Polyakov74}%
	\BibitemOpen
	\bibfield  {author} {\bibinfo {author} {\bibfnamefont {A.~M.}\ \bibnamefont
			{Polyakov}},\ }\bibfield  {title} {\bibinfo {title} {Particle spectrum in the
			quantum field theory},\ }\href@noop {} {\bibfield  {journal} {\bibinfo
			{journal} {Zh. Eksp. Teor. Fiz. Pis. Red.}\ }\textbf {\bibinfo {volume}
			{20}},\ \bibinfo {pages} {430} (\bibinfo {year} {1974})}\BibitemShut
	{NoStop}%
	\bibitem [{\citenamefont {Belavin}\ \emph {et~al.}(1975)\citenamefont
		{Belavin}, \citenamefont {Polyakov}, \citenamefont {Schwartz},\ and\
		\citenamefont {Tyupkin}}]{Belavin1975}%
	\BibitemOpen
	\bibfield  {author} {\bibinfo {author} {\bibfnamefont {A.~A.}\ \bibnamefont
			{Belavin}}, \bibinfo {author} {\bibfnamefont {A.~M.}\ \bibnamefont
			{Polyakov}}, \bibinfo {author} {\bibfnamefont {A.~S.}\ \bibnamefont
			{Schwartz}},\ and\ \bibinfo {author} {\bibfnamefont {Y.~S.}\ \bibnamefont
			{Tyupkin}},\ }\bibfield  {title} {\bibinfo {title} {{Pseudoparticle solutions
				of the {Yang-Mills} equations}},\ }\href
	{https://doi.org/10.1016/0370-2693(75)90163-X} {\bibfield  {journal}
		{\bibinfo  {journal} {Phys. Lett. B}\ }\textbf {\bibinfo {volume} {59}},\
		\bibinfo {pages} {85} (\bibinfo {year} {1975})}\BibitemShut {NoStop}%
	\bibitem [{\citenamefont {Coleman}(1985)}]{Coleman1985}%
	\BibitemOpen
	\bibfield  {author} {\bibinfo {author} {\bibfnamefont {S.}~\bibnamefont
			{Coleman}},\ }\href {https://doi.org/10.1017/CBO9780511565045} {\emph
		{\bibinfo {title} {Aspects of Symmetry: Selected Erice Lectures}}}\ (\bibinfo
	{publisher} {Cambridge University Press},\ \bibinfo {year}
	{1985})\BibitemShut {NoStop}%
	\bibitem [{\citenamefont {'t~Hooft}(1976)}]{Hooft76symmetry}%
	\BibitemOpen
	\bibfield  {author} {\bibinfo {author} {\bibfnamefont {G.}~\bibnamefont
			{'t~Hooft}},\ }\bibfield  {title} {\bibinfo {title} {Symmetry breaking
			through {Bell-Jackiw} anomalies},\ }\href
	{https://doi.org/10.1103/PhysRevLett.37.8} {\bibfield  {journal} {\bibinfo
			{journal} {Phys. Rev. Lett.}\ }\textbf {\bibinfo {volume} {37}},\ \bibinfo
		{pages} {8} (\bibinfo {year} {1976})}\BibitemShut {NoStop}%
	\bibitem [{\citenamefont {Derrick}(1964)}]{Derrick1964comments}%
	\BibitemOpen
	\bibfield  {author} {\bibinfo {author} {\bibfnamefont {G.~H.}\ \bibnamefont
			{Derrick}},\ }\bibfield  {title} {\bibinfo {title} {Comments on nonlinear
			wave equations as models for elementary particles},\ }\href@noop {}
	{\bibfield  {journal} {\bibinfo  {journal} {J. Math. Phys.}\ }\textbf
		{\bibinfo {volume} {5}},\ \bibinfo {pages} {1252} (\bibinfo {year}
		{1964})}\BibitemShut {NoStop}%
	\bibitem [{\citenamefont {Arbey}\ and\ \citenamefont
		{Mahmoudi}(2021)}]{Arbey2021}%
	\BibitemOpen
	\bibfield  {author} {\bibinfo {author} {\bibfnamefont {A.}~\bibnamefont
			{Arbey}}\ and\ \bibinfo {author} {\bibfnamefont {F.}~\bibnamefont
			{Mahmoudi}},\ }\bibfield  {title} {\bibinfo {title} {Dark matter and the
			early universe: A review},\ }\href
	{https://doi.org/https://doi.org/10.1016/j.ppnp.2021.103865} {\bibfield
		{journal} {\bibinfo  {journal} {Prog. Part. Nucl. Phys.}\ }\textbf {\bibinfo
			{volume} {119}},\ \bibinfo {pages} {103865} (\bibinfo {year}
		{2021})}\BibitemShut {NoStop}%
	\bibitem [{\citenamefont {Ginzburg}\ and\ \citenamefont
		{Landau}(1950)}]{Ginzburg50}%
	\BibitemOpen
	\bibfield  {author} {\bibinfo {author} {\bibfnamefont {V.~L.}\ \bibnamefont
			{Ginzburg}}\ and\ \bibinfo {author} {\bibfnamefont {L.~D.}\ \bibnamefont
			{Landau}},\ }\bibfield  {title} {\bibinfo {title} {On the theory of
			superconductivity},\ }\href@noop {} {\bibfield  {journal} {\bibinfo
			{journal} {Zh. Eksp. Teor. Fiz.}\ }\textbf {\bibinfo {volume} {20}},\
		\bibinfo {pages} {1064} (\bibinfo {year} {1950})}\BibitemShut {NoStop}%
	\bibitem [{\citenamefont {Bardeen}\ \emph {et~al.}(1957)\citenamefont
		{Bardeen}, \citenamefont {Cooper},\ and\ \citenamefont {Schrieffer}}]{BCS57}%
	\BibitemOpen
	\bibfield  {author} {\bibinfo {author} {\bibfnamefont {J.}~\bibnamefont
			{Bardeen}}, \bibinfo {author} {\bibfnamefont {L.~N.}\ \bibnamefont
			{Cooper}},\ and\ \bibinfo {author} {\bibfnamefont {J.~R.}\ \bibnamefont
			{Schrieffer}},\ }\bibfield  {title} {\bibinfo {title} {Microscopic theory of
			superconductivity},\ }\href {https://doi.org/10.1103/PhysRev.106.162}
	{\bibfield  {journal} {\bibinfo  {journal} {Phys. Rev.}\ }\textbf {\bibinfo
			{volume} {106}},\ \bibinfo {pages} {162} (\bibinfo {year}
		{1957})}\BibitemShut {NoStop}%
	\bibitem [{\citenamefont {Higgs}(1964)}]{Higgs64}%
	\BibitemOpen
	\bibfield  {author} {\bibinfo {author} {\bibfnamefont {P.~W.}\ \bibnamefont
			{Higgs}},\ }\bibfield  {title} {\bibinfo {title} {Broken symmetries and the
			masses of gauge bosons},\ }\href {https://doi.org/10.1103/PhysRevLett.13.508}
	{\bibfield  {journal} {\bibinfo  {journal} {Phys. Rev. Lett.}\ }\textbf
		{\bibinfo {volume} {13}},\ \bibinfo {pages} {508} (\bibinfo {year}
		{1964})}\BibitemShut {NoStop}%
	\bibitem [{\citenamefont {Altomare}\ and\ \citenamefont
		{Chang}(2013)}]{Altomare2013one}%
	\BibitemOpen
	\bibfield  {author} {\bibinfo {author} {\bibfnamefont {F.}~\bibnamefont
			{Altomare}}\ and\ \bibinfo {author} {\bibfnamefont {A.~M.}\ \bibnamefont
			{Chang}},\ }\href@noop {} {\emph {\bibinfo {title} {One-dimensional
				superconductivity in nanowires}}}\ (\bibinfo  {publisher} {John Wiley \&
		Sons},\ \bibinfo {year} {2013})\BibitemShut {NoStop}%
	\bibitem [{\citenamefont {Dashen}\ \emph
		{et~al.}(1974{\natexlab{b}})\citenamefont {Dashen}, \citenamefont
		{Hasslacher},\ and\ \citenamefont {Neveu}}]{Dashen74nonabelian}%
	\BibitemOpen
	\bibfield  {author} {\bibinfo {author} {\bibfnamefont {R.~F.}\ \bibnamefont
			{Dashen}}, \bibinfo {author} {\bibfnamefont {B.}~\bibnamefont {Hasslacher}},\
		and\ \bibinfo {author} {\bibfnamefont {A.}~\bibnamefont {Neveu}},\ }\bibfield
	{title} {\bibinfo {title} {{Nonperturbative methods and extended-hadron
				models in field theory. III. Four-dimensional non-Abelian models}},\ }\href
	{https://doi.org/10.1103/PhysRevD.10.4138} {\bibfield  {journal} {\bibinfo
			{journal} {Phys. Rev. D}\ }\textbf {\bibinfo {volume} {10}},\ \bibinfo
		{pages} {4138} (\bibinfo {year} {1974}{\natexlab{b}})}\BibitemShut {NoStop}%
	\bibitem [{\citenamefont {Witzel}(2019)}]{Witzel2019review}%
	\BibitemOpen
	\bibfield  {author} {\bibinfo {author} {\bibfnamefont {O.}~\bibnamefont
			{Witzel}},\ }\bibfield  {title} {\bibinfo {title} {Review on composite
			{Higgs} models},\ }\href {https://doi.org/10.22323/1.334.0006} {\bibfield
		{journal} {\bibinfo  {journal} {Proc. Sci.}\ }\textbf {\bibinfo {volume}
			{334}},\ \bibinfo {pages} {006} (\bibinfo {year} {2019})}\BibitemShut
	{NoStop}%
	\bibitem [{\citenamefont {{Particle Data Group}}\ \emph
		{et~al.}(2020)\citenamefont {{Particle Data Group}}, \citenamefont {Zyla},
		\citenamefont {Barnett}, \citenamefont {Beringer}, \citenamefont {Dahl},
		\citenamefont {Dwyer}, \citenamefont {Groom}, \citenamefont {Lin},
		\citenamefont {Lugovsky}, \citenamefont {Pianori} \emph
		{et~al.}}]{Particle2020review}%
	\BibitemOpen
	\bibfield  {author} {\bibinfo {author} {\bibnamefont {{Particle Data
					Group}}}, \bibinfo {author} {\bibfnamefont {P.}~\bibnamefont {Zyla}},
		\bibinfo {author} {\bibfnamefont {R.}~\bibnamefont {Barnett}}, \bibinfo
		{author} {\bibfnamefont {J.}~\bibnamefont {Beringer}}, \bibinfo {author}
		{\bibfnamefont {O.}~\bibnamefont {Dahl}}, \bibinfo {author} {\bibfnamefont
			{D.}~\bibnamefont {Dwyer}}, \bibinfo {author} {\bibfnamefont
			{D.}~\bibnamefont {Groom}}, \bibinfo {author} {\bibfnamefont {C.-J.}\
			\bibnamefont {Lin}}, \bibinfo {author} {\bibfnamefont {K.}~\bibnamefont
			{Lugovsky}}, \bibinfo {author} {\bibfnamefont {E.}~\bibnamefont {Pianori}},
		\emph {et~al.},\ }\bibfield  {title} {\bibinfo {title} {Review of particle
			physics},\ }\href@noop {} {\bibfield  {journal} {\bibinfo  {journal} {Prog.
				Theor. Exp. Phys.}\ }\textbf {\bibinfo {volume} {2020}},\ \bibinfo {pages}
		{083C01} (\bibinfo {year} {2020})}\BibitemShut {NoStop}%
	\bibitem [{\citenamefont {Choppin}\ \emph {et~al.}(2013)\citenamefont
		{Choppin}, \citenamefont {Liljenzin}, \citenamefont {Rydberg},\ and\
		\citenamefont {Ekberg}}]{Choppin2002radiochemistry}%
	\BibitemOpen
	\bibfield  {author} {\bibinfo {author} {\bibfnamefont {G.}~\bibnamefont
			{Choppin}}, \bibinfo {author} {\bibfnamefont {J.-O.}\ \bibnamefont
			{Liljenzin}}, \bibinfo {author} {\bibfnamefont {J.}~\bibnamefont {Rydberg}},\
		and\ \bibinfo {author} {\bibfnamefont {C.}~\bibnamefont {Ekberg}},\ }\href
	{https://doi.org/https://doi.org/10.1016/C2011-0-07260-5} {\emph {\bibinfo
			{title} {Radiochemistry and nuclear chemistry}}},\ \bibinfo {edition} {4th}\
	ed.\ (\bibinfo  {publisher} {Elsevier},\ \bibinfo {year} {2013})\BibitemShut
	{NoStop}%
	\bibitem [{\citenamefont {Lindl}(1995)}]{Lindl1995development}%
	\BibitemOpen
	\bibfield  {author} {\bibinfo {author} {\bibfnamefont {J.}~\bibnamefont
			{Lindl}},\ }\bibfield  {title} {\bibinfo {title} {Development of the
			indirect-drive approach to inertial confinement fusion and the target physics
			basis for ignition and gain},\ }\href@noop {} {\bibfield  {journal} {\bibinfo
			{journal} {Phys. plasmas}\ }\textbf {\bibinfo {volume} {2}},\ \bibinfo
		{pages} {3933} (\bibinfo {year} {1995})}\BibitemShut {NoStop}%
	\bibitem [{\citenamefont {Bethe}(1950)}]{Bethe1950hydrogen}%
	\BibitemOpen
	\bibfield  {author} {\bibinfo {author} {\bibfnamefont {H.~A.}\ \bibnamefont
			{Bethe}},\ }\bibfield  {title} {\bibinfo {title} {The hydrogen bomb},\
	}\href@noop {} {\bibfield  {journal} {\bibinfo  {journal} {B. Atom. Sci.}\
		}\textbf {\bibinfo {volume} {6}},\ \bibinfo {pages} {99} (\bibinfo {year}
		{1950})}\BibitemShut {NoStop}%
	\bibitem [{\citenamefont {Norton}(2002)}]{Norton2002introduction}%
	\BibitemOpen
	\bibfield  {author} {\bibinfo {author} {\bibfnamefont {A.}~\bibnamefont
			{Norton}},\ }\href@noop {} {\emph {\bibinfo {title} {An Introduction to
				Astrophysics}}}\ (\bibinfo  {publisher} {The Open University},\ \bibinfo
	{year} {2002})\BibitemShut {NoStop}%
	\bibitem [{\citenamefont {Revaz}\ and\ \citenamefont
		{Jablonka}(2018)}]{Revaz2018pushing}%
	\BibitemOpen
	\bibfield  {author} {\bibinfo {author} {\bibfnamefont {Y.}~\bibnamefont
			{Revaz}}\ and\ \bibinfo {author} {\bibfnamefont {P.}~\bibnamefont
			{Jablonka}},\ }\bibfield  {title} {\bibinfo {title} {{Pushing back the
				limits: detailed properties of dwarf galaxies in a $\Lambda$CDM universe}},\
	}\href@noop {} {\bibfield  {journal} {\bibinfo  {journal} {Astron.
				Astrophys.}\ }\textbf {\bibinfo {volume} {616}},\ \bibinfo {pages} {A96}
		(\bibinfo {year} {2018})}\BibitemShut {NoStop}%
	\bibitem [{\citenamefont {Posti}\ and\ \citenamefont
		{Helmi}(2019)}]{Posti2019mass}%
	\BibitemOpen
	\bibfield  {author} {\bibinfo {author} {\bibfnamefont {L.}~\bibnamefont
			{Posti}}\ and\ \bibinfo {author} {\bibfnamefont {A.}~\bibnamefont {Helmi}},\
	}\bibfield  {title} {\bibinfo {title} {{Mass and shape of the Milky Wayâ€™s
				dark matter halo with globular clusters from Gaia and Hubble}},\ }\href@noop
	{} {\bibfield  {journal} {\bibinfo  {journal} {Astron. Astrophys.}\ }\textbf
		{\bibinfo {volume} {621}},\ \bibinfo {pages} {A56} (\bibinfo {year}
		{2019})}\BibitemShut {NoStop}%
	\bibitem [{\citenamefont {Shi}(2019)}]{Shi2019force}%
	\BibitemOpen
	\bibfield  {author} {\bibinfo {author} {\bibfnamefont {Y.}~\bibnamefont
			{Shi}},\ }\bibfield  {title} {\bibinfo {title} {Force, curvature, or mass:
			disambiguating causes of uniform gravity},\ }\href@noop {} {\bibfield
		{journal} {\bibinfo  {journal} {arXiv:1908.02159}\ } (\bibinfo {year}
		{2019})}\BibitemShut {NoStop}%
\end{thebibliography}

%

\end{document}